A spectrum of routing strategies for brain networks

Andrea Avena-Koenigsberger[1*], Xiaoran Yan[2], Artemy Kolchinsky[3], Martijn van den Heuvel [4], Patric Hagmann[5,6], Olaf Sporns[1,2]

1. Department of Psychological and Brain Sciences, Indiana University, Bloomington, IN, USA
2. IU Network Institute, Indiana University, Bloomington, IN, USA
3. Santa Fe Institute, Santa Fe, NM, USA
4. Brain Center Rudolf Magnus, University Medical Center Utrecht, University Utrecht, The Netherlands
5. Department of Radiology, Centre Hospitalier Universitaire Vaudois (CHUV) and University of Lausanne (UNIL), Lausanne, Switzerland
6. Signal Processing Laboratory 5 (LTS5), Ecole Polytechnique Federale de Lausanne (EPFL), Lausanne, Switzerland

* Corresponding Author
Andrea Avena-Koenigsberger
aiavenak@indiana.edu

**Abstract**

Communication of signals among nodes in a complex network poses fundamental problems of efficiency and cost. Routing of messages along shortest paths requires global information about the topology, while spreading by diffusion, which operates according to local topological features, is informationally "cheap" but inefficient. We introduce a stochastic model for network communication that combines varying amounts of local and global information about the network topology. The model generates a continuous spectrum of dynamics that converge onto shortest-path and random-walk (diffusion) communication processes at the limiting extremes. We implement the model on two cohorts of human connectome networks and investigate the effects of varying amounts of local and global information on the network's communication cost. We identify routing strategies that approach a (highly efficient) shortest-path communication process with a relatively small amount of global information. Moreover, we show that the cost of routing messages from and to hub nodes varies as a function of the amount of global information driving the system's dynamics. Finally, we implement the model to identify individual subject differences from a communication dynamics point of view. The present framework departs from the classical shortest paths vs. diffusion dichotomy, suggesting instead that brain networks may exhibit different types of communication dynamics depending on varying functional demands and the availability of resources.

**Introduction**

The function of many real world complex networks is to relay information within and between their constituent elements. Efficient communication, i.e. the passing of information at high speed and high reliability at low cost to the system, is essential to the functioning of systems in many domains, ranging from technological to social and biological applications. For example, communication is central to the operation of brain networks, as it is necessary for information integration and for distributed neural computation [1]. However, the mechanisms that enable information to flow efficiently among large numbers of distributed elements interacting through a complex topology remain mostly unexplained.

Previous work on optimal routing in networks highlighted the importance of small-world topologies for promoting short communication pathways at low wiring cost [2,3]. Indeed, information transfer that takes place through topologically shortest paths is both fast and direct, and reduces a message's vulnerability to errors and attack [4]. Yet, such a communication model also has disadvantages: it discounts the vast majority of a network's structural connections [5,6], it is prone to bottlenecks and congestion [7-9], and it lacks robustness to edge failures [10]. Most importantly, a system's ability to route along shortest paths relies on all of the system's elements having information about the global topology of the network [11,12]. Therefore, an explicit analysis of the costs and benefits of efficient communication should take into account the cost associated with global information, in addition to better-known costs such as wiring and energy consumption [1,13-17]. We refer to the cost of the information necessary for signal routing as the *informational cost*.

A drastically different picture emerges if we discard the premise that the system's elements are capable of accessing information about the global topology of the network. Under this scenario, signals are dispersed according to a random walk or diffusion process [18-20], driven only by local topological properties. While diffusion has no associated cost of storing global topological information, communication is inefficient if measured in terms of the time needed for a signal to arrive at a specific destination. This results in increased vulnerability to signal corruption and slower integration of information as signals are broadcast and spread indiscriminately across the network.

While shortest paths and diffusion have been extensively studied in the context of network communication, they merely represent the extremes of a spectrum of communication processes that deserve greater attention. As an example, for some types of network topologies, a preferential choice policy where messages are preferentially routed towards high degree nodes [21, 22] decreases search times significantly compared to random walks, yet the informational cost is small since nodes only need to "know" the degree of their neighbors. Brain networks are a case in point: on average, shortest paths tend to follow a low-to-high and then high-to-low degree sequence [23] and closeness centrality sequence [24], suggesting that efficient routing patterns in brain networks could be driven by a mixture of degree and closeness preferential choice policies. Preferential policies are often modeled as biased random walks [25], where the motion of a random walker located at a given node is biased according to an attribute (e.g. degree) associated with the neighboring nodes. It has been shown that biased random walks can generate relatively efficient communication processes (high speed, low cost) and are able to account for navigation rules that are observed in real world systems [26-29], offering alternative interpretations of node centralities and community structures [30].

Here, we focus on a specific family of biased random walks, governed by routing strategies generated by a stochastic model that combines local and global information about the network topology. This framework allows us to explore a continuous spectrum of dynamics that converge onto shortest-path communication processes at one extreme, and random-walk (diffusion) communication processes at the other extreme. We apply this framework to investigate *communication cost* from a dynamic point of view in large-scale brain networks. We suggest that brain networks may exhibit different types of communication dynamics depending on varying functional demands and the availability of resources.

**A continuous spectrum of routing strategies combining local and global information.**
We model messages or signals transferred from a source brain region to a target brain region as random walkers traversing a brain network, where network nodes and edges represent small cortical parcels that are connected by bundles of axons. We consider the dynamics of such random walkers (signals/messages) on the network, where walkers must reach an a priori specified target node *t*. Formally, let **X** be a random variable indicating the current node of the walker, **Y** the random variable indicating the node to which the walker will move in the next time step, and **T** the random variable indicating the target node where the walk will terminate (we assume that all nodes can be reached from all nodes in finite time). For all *t*, we denote the transition probabilities at **X**=*i* as:

$P^t_{ij} = Pr(\mathbf{Y} = j|\mathbf{X} = i, \mathbf{T} = t)$ where $\sum_j P^t_{ij} = 1$, and $P^t_{ij} = 0$ when there is no connection between nodes $i$ and $j$. Finally, the walk ends when $i = t$, in which case $P^t_{ij} = 1$ for $j=t$ and 0 for all other $j$. Formally, the network dynamics for each separate target $t$ form a Markov chain with state $t$ as an absorbing state (see Methods). The set of transition probabilities for all $t$ express the routing strategy that governs the dynamics of walkers (signals) navigating the network.

We specify transition probabilities at every node using a family of dynamical processes that combine local and global information about the network's topology. To this end, we define the dynamics of the system by controlling the effect of global information using the following stochastic model:

$$P_\lambda(Y = j \mid X = i, T = t) = \exp(-(\lambda(d_{ij} + g_{jt}) + d_{ij}))\frac{1}{Z^t_i}$$

where $Z^t_i = \sum_j \exp(-(\lambda(d_{ij} + g_{jt}) + d_{ij}))$ is a normalization factor. Transition probabilities are governed by two sources of information:

- a *local* source of information $d_{ij}$ denoting the length of the edge connecting $i$ and $j$ ( $d_{ij} \neq \infty$ if and only if a connection between $i$ and $j$ exists).
- a *global* source of information $d_{ij} + g_{jt}$, denoting the minimum distance from node $i$ to target $t$ though node $j$. This is the sum of the distance between node $i$ and neighbor node $j$ ($d_{ij}$), plus the distance from node $j$ to the target node $t$ through the shortest path ($g_{jt}$ - note that this term has no dependence on $i$).

The parameter $\lambda$ controls the extent to which transition probabilities are shaped by global information. Most importantly, $\lambda$ gradually changes the dynamics on the network from an unbiased random walk towards a shortest-path routing strategy (see Fig 1):

- When $\lambda=0$, a walker's motion is driven only by local information. Transition probabilities are simply given by $P_\lambda(Y = j \mid X = i, T = t) = \exp(-d_{ij})\frac{1}{Z^t_i}$ and do not depend on the target node (nonetheless, the walk still terminates when it eventually reaches the target node $t$). In the case of brain networks, where edge-weights express connection strengths or node proximities in the interval (0,1) (this can always be enforced through a unique linear normalization function [32]; See Methods), we apply the proximity-to-distance function $d_{ij} = -\log(w_{ij})$ and map all edge-weights onto edge-distances. The resulting dynamics, $P_0(Y = j \mid X = i, T = t) = w_{ij} \frac{1}{s_i}$, where $s_i = \sum_j w_{ij}$ is the strength of node $i$, defines an unbiased random walk on the network where walkers favor transitions through edges with shorter connection distances (i.e. closer proximities). We refer to the unbiased random walk as the *reference navigation strategy*, $P^{ref}$, as it represents a null model of navigation that would naturally take place on the network if no bias is introduced.

- When $\lambda \to \infty$ global information governs the model and transition probabilities converge to $P^t_{ij} = 1$ if the edge $\{i, j\}$ lies on the (unique) shortest path between $i$ and $t$ (degenerate shortest paths, i.e. more than one shortest path from $i$ to $t$, are unlikely in weighted networks, but see Methods for the case where degenerative shortest paths exist) and $P^t_{ij} = 0$ otherwise. Hence, statistics computed on such walks will correspond to a "shortest-path" routing strategy - in particular average walk length will be equal to shortest path length.

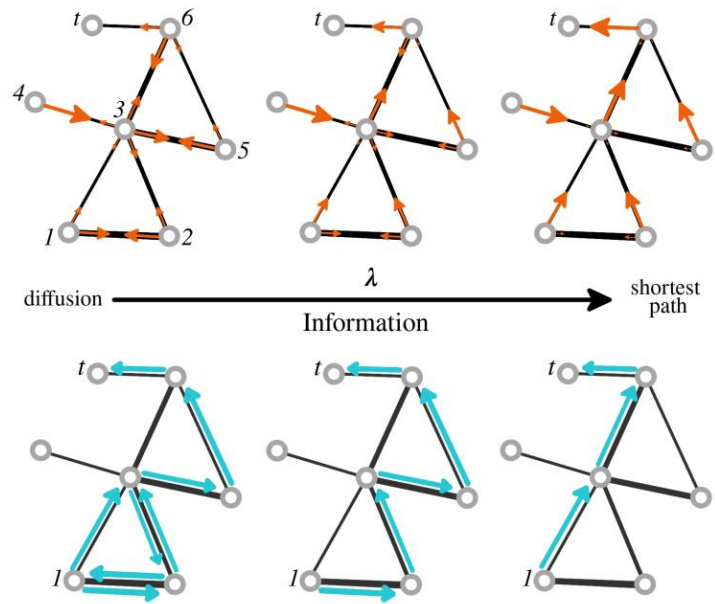

Fig 1. A spectrum of routing strategies. The parameter $\lambda$ controls the extent to which routing strategies (transition probabilities) are reshaped by global information. Toy networks in the top row illustrate how transition probabilities, represented by orange arrows, are reshaped as the parameter $\lambda$ increases. At each node, the orange arrows are proportional to the probability of a walker moving to a neighboring node via that edge. Blue arrows on the toy networks in the bottom row illustrate a possible walk followed by a random walker (signal) going from node *1* to node *t*, while operating according to the routing strategy represented by the orange arrows. When λ=0, transition probabilities at each node are proportional to the strength of its connections. Random walkers operating under this routing strategy (the *reference navigation strategy*, $P^{ref}$) diffuse through the network until they eventually arrive at the target node. When $\lambda \to \infty$, transition probabilities at each node route walkers through the shortest path to the target node; a walker starting at node 1 will follow the shortest path to node t, as illustrated by the blue arrows. In the middle of the spectrum, walker's dynamics are influenced by global information but still driven partially by local topological properties. Notice that only transition probabilities vary with λ while the underlying network structure remains invariant.

It is worth noting that the model acts on the routing strategies by changing the transition probabilities at each node, but the topology and weight structure of the network remain unchanged (see Fig 1).

**The cost of reshaping the system's dynamics.**

We are interested in characterizing the *communication cost* of the dynamics generated by our model as we gradually increase λ. Here, we focus on two aspects of the cost associated with a communication process. First, we consider a *transmission cost*, which is the cost associated with messages being transmitted from one node to another. Second, we consider an *informational cost*, which is the cost associated with using global information to reshape the system's dynamics and thus route messages efficiently.

We consider a walker navigating the network and acting according to the routing strategies $P_\lambda(\mathbf{Y}|\mathbf{X},\mathbf{T})$. Let $c_\lambda^{trans}(i,t) = \sum_j P_\lambda(\mathbf{Y}=j|\mathbf{X}=i,\mathbf{T}=t) d_{ij}$ be the *immediate transmission cost* at node *i* for a walker going to node *t* with routing strategy $P_\lambda(\mathbf{Y}|\mathbf{X}=i,\mathbf{T}=t)$. The *immediate transmission cost* quantifies the cost associated with $\mathbf{X}=i$ partaking in the communication process by relaying the message to one of its neighbors, and in this setting it is equivalent to the expected distance that a walker at node *i* has to travel to move to a neighbor of *i*. Let $n^t_\lambda(i,k)$ be the mean number of times node *k* is visited by a walker starting at a source node $\mathbf{X_0}=i$ and acting according to a routing strategy $P_\lambda(\mathbf{Y}|\mathbf{X}=i,\mathbf{T}=t)$. We define the transmission cost of a walk starting at source node $\mathbf{X_0}=i$ and terminating at the target node *t* as the sum of the *immediate transmission costs* accumulated at each visited node along a walk, that is $C_\lambda^{trans}(i,t) = \sum_k n^t_\lambda(i,k) c_\lambda^{trans}(k,t)$. Thus, a walk's transmission cost is equivalent to the mean walk length between nodes *i* and *t*, under the routing strategy defined by λ. Noting that the transmission cost is not a symmetric measure, (i.e. $C_\lambda^{trans}(i,t)$ may not be the same as $C_\lambda^{trans}(t,i)$, except for when $\lambda \to \infty$), we can

define the average transmission cost of a node acting as a source as $\vec{C}_\lambda^{trans}(i) = \frac{1}{N}\sum_t c_\lambda^{trans}(i,t)$, and the average transmission cost of a node acting as a target as $\overleftarrow{C}_\lambda^{trans}(t) = \frac{1}{N}\sum_i c_\lambda^{trans}(i,t)$. These measures quantify the *source and target closeness centrality* of each node under a routing strategy: $\vec{C}_\lambda^{trans}(i)$ quantifies the average walk length from a node *i* to any other target node in the network, whereas $\overleftarrow{C}_\lambda^{trans}(t)$ quantifies the average walk length from any source node to the target node *t*.

To quantify the informational cost associated with routing messages to a target node *t* under the routing strategy P$_\lambda$(**Y**| **X**=*i*,**T** = *t*), we define $c_\lambda^{info}(i,t)$ = KL(P$_\lambda$(**Y**| **X**=*i*,**T** = *t*)||$P^{ref}$(**Y**| **X**=*i*,**T** = *t*)) as the *informational cost* at node **X**=*i*, measuring the Kullback-Leibler divergence between the routing strategy P$_\lambda$(**Y**|*i*,*t*) and the reference routing strategy $P^{ref}$ (**Y**|*i*,*t*). The *informational cost* quantifies the effect of the bias due to global information by measuring how much reshaping of the system's dynamics has taken place at node **X**=*i* [33]. Then, the informational cost of routing a message from a starting at node **X**=*i* to a target node *t* is the weighted average *informational cost* across all nodes in the network, weighted by the frequency with which each node is visited along the walk: $C_\lambda^{info}(i,t) = \sum_k \left(\frac{n_\lambda^t(i,k)}{\sum_k n_\lambda^t(i,k)} c_\lambda^{trans}(k,t)\right)$.

Finally, we can define the average informational cost of a node acting as a source as $\vec{C}_\lambda^{info}(i) = \frac{1}{N}\sum_t c_\lambda^{info}(i,t)$, and the average informational cost of a node acting as a target as $\overleftarrow{C}_\lambda^{info}(t) = \frac{1}{N}\sum_i c_\lambda^{info}(i,t)$.

In the following sections we will study the communication costs of routing strategies generated by our stochastic model applied to the structural brain connectivity matrices of two cohorts of healthy subjects. In the main text, we focus on 173 unrelated subjects from the Human Connectome Project (HCP) dataset [34,35]. The Supplementary Fig S7 - S11 show results from the replication dataset (LAU), composed of 40 healthy subjects (see Methods). We first analyze cost measures at the global, nodal, and pairwise level, measured and averaged across all subjects (within each cohort). In the last section we demonstrate the utility of this approach for identifying individual subject differences from a communication dynamics point of view, hence departing from the classical routing vs. diffusion dichotomy [12,36].

**Brain networks are more efficient within an intermediate region of the communication spectrum.**
By construction, the transmission and informational cost have a competing relationship (or trade-off) such that as we increase λ in the stochastic model, the mean walk lengths ($C_\lambda^{trans}$) of messages acting according to P$_\lambda$ become shorter while the bias effect due to global information ($C_\lambda^{info}$) increases. This trade-off is shown in Fig 2a where averages of $C_\lambda^{trans}$ and $C_\lambda^{info}$ across all {$i,t$} pairs are plotted as a function of λ. It can be seen that $C_\lambda^{trans}$, measuring the average walk length, approaches a shortest path-length regime at around λ = 1 (ln(λ) = 0 in Fig 2), suggesting that in this regime messages can be efficiently routed at a low informational cost.

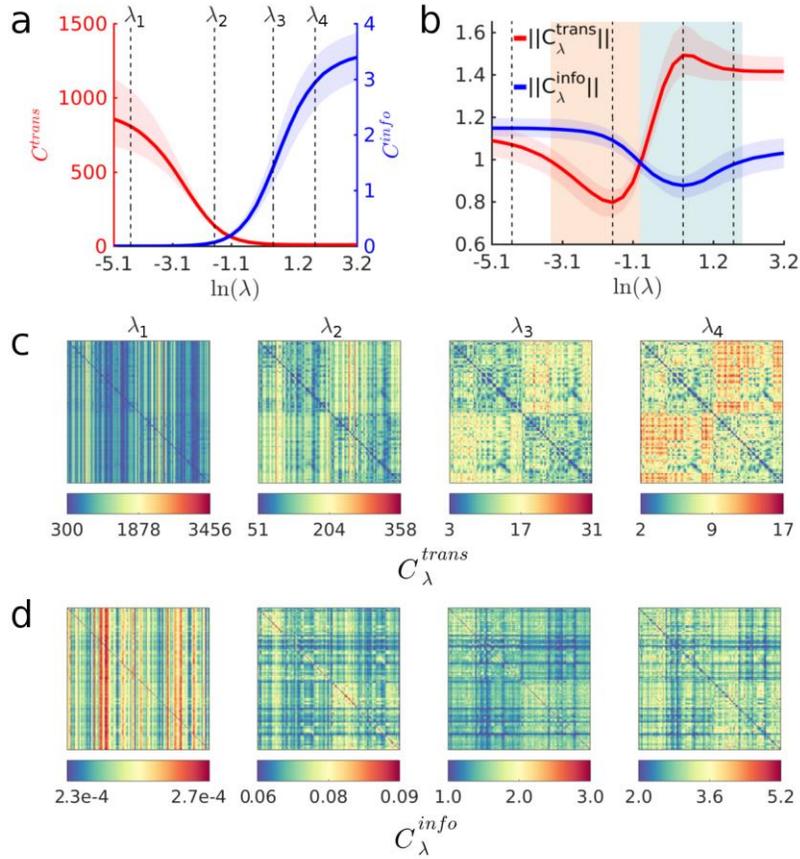

Fig 2: A spectrum of communication processes. (a) Averages of $C_\lambda^{trans}$ (red) and $C_\lambda^{info}$ (blue) across all node pairs, as a function of λ. Solid red and blue lines correspond to the median across all subjects, whereas the shaded red and blue regions denote the 95$^{th}$ percentile. (b) Averages of $\|C_\lambda^{trans}\|$ (red) and $\|C_\lambda^{info}\|$ (blue) across all node pairs. These curves are computed by normalizing $C_\lambda^{trans}$ and $C_\lambda^{info}$ with respect to the same cost measures computed on ensembles of 500 randomized networks (per subject). Shaded red and blue areas indicate sections of the curves $\|C_\lambda^{trans}\|$ and $\|C_\lambda^{info}\|$ that are smaller than 1, respectively, indicating the regions in the spectrum where the communication cost of empirical networks is smaller than the cost computed on the randomized ensembles. The dashed vertical lines are placed at the minimum and maximum of $\|C_\lambda^{trans}\|$ ($\lambda_2$ and $\lambda_3$, respectively), and at two points near the extremes of the spectrum (($\lambda_1$ and $\lambda_4$). (c) pairwise values of $C_\lambda^{trans}(i,t)$ for all node pairs. (d) pairwise values of $C_\lambda^{info}(i,t)$ for all node pairs. In all panels, $\lambda_1=e^{-4.49}$, $\lambda_2=e^{-1.64}$, $\lambda_3=e^{0.37}$ and $\lambda_4=e^{1.79}$.

Next, we consider an ensemble of random networks and compare average transmission and informational costs incurred in empirical brain networks and in randomized ensembles of networks. All randomized networks preserve node degree, node strength (evaluated with respect to the proximity edge-weights), and the network's weight distribution (see Methods). We generate routing strategies $P_\lambda$ for all randomized networks and normalize the cost measures $C_\lambda^{trans}$ and $C_\lambda^{info}$ of each subject's empirical brain network with respect to the average cost measures computed across the corresponding randomized counterparts. Fig 2b shows normalized cost measures $\|C_\lambda^{trans}\| = C_\lambda^{trans}(emp)/C_\lambda^{trans}(rand)$ (red line) and $\|C_\lambda^{info}\| = C_\lambda^{info}(emp)/C_\lambda^{info}(rand)$ (blue line) as a function of $\lambda$. In accordance with prior work (37-39), we find that average walk lengths are shorter for random networks (i.e. $\|C_\lambda^{trans}\| > 1$) at the extremes of the spectrum, representing the unbiased random walk ($P^{ref}$) and shortest path regimes. Interestingly, our analysis reveals an interval of $\lambda$ values (shaded region in Fig 2b) for which empirical networks exhibit shorter walk-lengths compared to the randomized counterparts (i.e. $\|C_\lambda^{trans}\| < 1$). Moreover, the informational cost behaves similarly, although the regions $\|C_\lambda^{info}\| < 1$ and $\|C_\lambda^{trans}\| < 1$ barely overlap. Overall, these results show that the randomized counterparts of empirical brain networks are more efficient only at the extremes of the communication spectrum.

Fig 2c and 2d show pairwise $C_\lambda^{trans}$ and $C_\lambda^{info}$ (median across subjects) computed for routing strategies generated with $\lambda_1=e^{-4.49}$, $\lambda_2=e^{-1.64}$, $\lambda_3=e^{0.37}$ and $\lambda_4=,e^{1.79}$ (see dashed vertical lines in Fig 2a and 2b). These values of $\lambda$ correspond to two points located near the extremes of the communication spectrum, and two points located at the minimum and maximum of the curve $\|C_\lambda^{trans}\|$, where the empirical networks are most and least efficient compared to their randomized counterparts. As evidenced by the column-like patterns in the matrices corresponding to $\lambda_1$ and $\lambda_2$, the dynamics of messages navigating the network are strongly determined by the local connectivity of the target node when the global information bias is small. As the bias increases and routing strategies depart from the reference strategy $P^{ref}$, the dynamics of messages are less dependent on the target node only. Finally, as walk-lengths converge towards shortest-path, the transmission cost becomes symmetric, i.e., $C_\lambda^{trans}(i,t) = C_\lambda^{trans}(t,i)$.

**Source vs. target communication cost**

We now analyze cost measures at the nodal level. Fig 3a and 3b show scatter plots of the average source and target transmission costs ($\vec{C}_\lambda^{trans}$ and $\overleftarrow{C}_\lambda^{trans}$, respectively), and the average source and target informational costs ($\vec{C}_\lambda^{info}$ and $\overleftarrow{C}_\lambda^{info}$, respectively) associated to all nodes (median across all subjects) for the same values of λ highlighted in Fig 2. Nodes are colored according to their membership in functional intrinsic connectivity networks (ICNs; see Methods), highlighting a tendency of some ICNs to contain an overabundance of costly sources and/or targets, while other ICNs' cost varies as a function of λ. For example, nodes belonging to the somatomotor network (SM, colored green) tend to exhibit a high $\vec{C}_\lambda^{trans}$ and low $\overleftarrow{C}_\lambda^{trans}$ for $\lambda < e^{0.37}$, while they also exhibit a consistent low $\vec{C}_\lambda^{info}$; nodes belonging to the visual network (VIS, colored red) exhibit high $\vec{C}_\lambda^{info}$ and $\overleftarrow{C}_\lambda^{info}$ for $\lambda > e^{-4.49}$, while $\vec{C}_\lambda^{trans}$ and $\overleftarrow{C}_\lambda^{trans}$ vary as a function of λ.

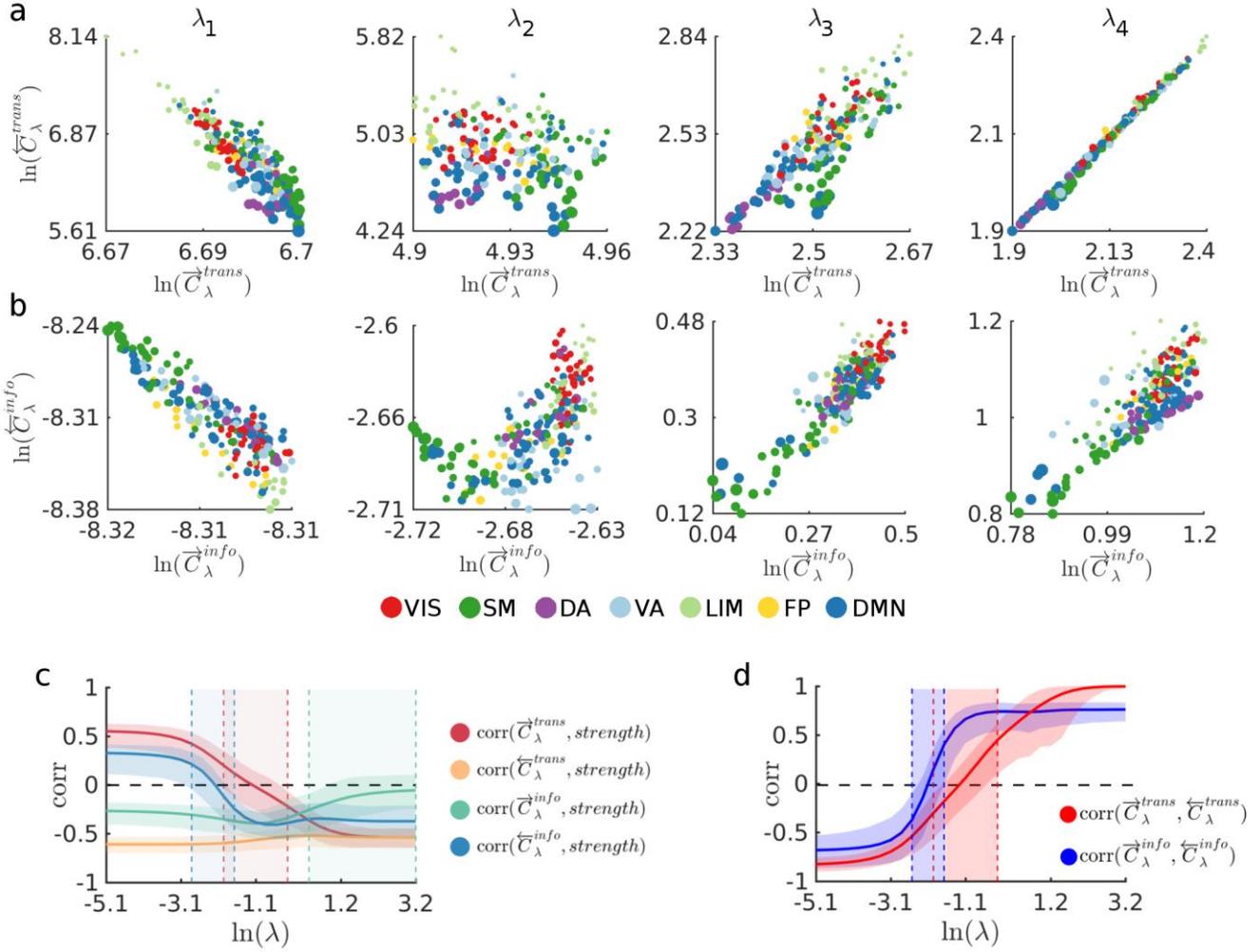

Fig 3: Nodal average transmission costs for four increasingly biased routing strategies. (a) Scatter plots show the transmission cost associated to each node when it acts as source ($\vec{C}_\lambda^{trans}$) and target ($\overleftarrow{C}_\lambda^{trans}$) during communication processes taking place under routing strategies generated with the values $\lambda_1$, $\lambda_2$, $\lambda_3$ and $\lambda_4$. (b) Scatter plots show the transmission cost associated to each node when it acts as source ($\vec{C}_\lambda^{info}$) and target ($\overleftarrow{C}_\lambda^{info}$) during communication processes taking place under routing strategies generated with the values $\lambda_1$, $\lambda_2$, $\lambda_3$ and $\lambda_4$. Markers in the scatter plots in (a) and (b), representing each node, are colored according to the node's membership in the 7 intrinsic connectivity networks (ICN) defined by Yeo et al. (2011) [61]: Visual (VIS), Somatomotor (SM), Dorsal Attention (DA), Ventral Attention (VA), Limbic (LIM), Frontal Parietal (FP), and Default Mode Network (DMN). The size of the markers is proportional to node's strength. (c) Correlations between node strength and $\vec{C}_\lambda^{trans}$ (red), $\overleftarrow{C}_\lambda^{trans}$ (orange), $\vec{C}_\lambda^{info}$ (green) and $\overleftarrow{C}_\lambda^{info}$ (blue) as a function of $\lambda$. Solid lines show median correlation across all subjects, shaded areas surrounding the lines show 95th percentile. Shaded colored areas between the vertical dashed lines indicate regions where the correlations were not significant (p > 0.001). (d) Correlation between $\vec{C}_\lambda^{trans}$ and $\overleftarrow{C}_\lambda^{trans}$ (red), and $\vec{C}_\lambda^{info}$ and $\overleftarrow{C}_\lambda^{info}$ (blue), as a function of $\lambda$. Solid lines show medians across all subjects and shaded areas surrounding solid lines show the 95th

percentile. Shaded areas between the vertical dashed lines indicate areas where correlation values were not significant (p > 0.001). In all panels, $\lambda_1=e^{-4.49}$, $\lambda_2=e^{-1.64}$, $\lambda_3=e^{0.37}$ and $\lambda_4=e^{1.79}$.

In order to assess to what extent high or low nodal costs are driven by the network's overall topology, as opposed to nodal degree or strength distribution, we standardize nodal costs with respect to the corresponding nodal cost distributions measured on the randomized network ensembles. Significantly high or low standardized nodal cost measures are indicative of global connectivity patterns that are encountered only in empirical brain networks. Supplementary Fig S1 - S4 show thresholded z-scores (*α = 0.01*) for all nodal cost measures as a function of lambda. As expected, near the extremes of the spectrum (λ = 0 and λ > 1), most nodes exhibit significantly higher costs, compared to the randomized networks, however, significantly low cost regions are found in the middle of the spectrum. Prominent low $\vec{C}_\lambda^{info}$ regions include the right and left hemisphere superior frontal and caudal middle frontal, precentral, paracentral and postcentral regions; low $\overleftarrow{C}_\lambda^{info}$ regions overlap with the low $\vec{C}_\lambda^{info}$ regions, but also include the right and left posterior cingulate, the supramarginal gyrus, the superior parietal cortex, the precuneus, and the inferior parietal cortex. Prominent low $\vec{C}_\lambda^{trans}$ regions are mainly located in the frontal cortex (frontal pole, medial orbital frontal and rostral middle frontal regions), right and left superior parietal regions, the right and left precuneus, and the left cuneus. No significantly low $\overleftarrow{C}_\lambda^{trans}$ regions were identified.

Our analyses also reveal a varying relationship (as a function of λ) between the nodal cost measures and node strength (see Fig 3c). At the extremes of the spectrum, transmission cost is strongly driven by node degree. When λ=0, the correlation between node strength and $\vec{C}_\lambda^{trans}$ and $\overleftarrow{C}_\lambda^{trans}$ is *r = 0.55* and *r = -0.61*, respectively (*p < 0.001*), indicating that high degree nodes (hubs) are costly sources but low cost targets with respect to transmission cost. In other words, when the global information bias is low (or zero), messages can be routed at a low transmission cost from any brain region to a hub; conversely, routing a message from a hub to any brain region incurs a high transmission cost. At the other end of the spectrum (i.e. for large values of λ), hub nodes are low cost sources and targets with respect to transmission cost (*r = -0.53, p<0.001* ; note that the orange and red lines in Fig 3c converge). However, in the middle of the spectrum, the average correlation between node strength and $\vec{C}_\lambda^{trans}$ is close to zero, whereas the correlation between node strength and $\overleftarrow{C}_\lambda^{trans}$ remains significant (*r ≈ -0.5 , p<0.001*) throughout the entire spectrum. The relationship between source and target costs also varies as a function of λ (see Fig 3d). For low values of λ, both $\vec{C}_\lambda^{trans}$ and $\overleftarrow{C}_\lambda^{trans}$, and $\vec{C}_\lambda^{info}$ and $\overleftarrow{C}_\lambda^{info}$ are negatively correlated. In other words, nodes that are costly sources are efficient targets, and nodes that are costly targets are efficient sources. However, the correlations undergo a sign flip as λ increases and $\vec{C}_\lambda^{trans}$ and $\overleftarrow{C}_\lambda^{trans}$, and $\vec{C}_\lambda^{info}$ and $\overleftarrow{C}_\lambda^{info}$ become positively correlated. Note that the correlation between $\vec{C}_\lambda^{trans}$ and $\overleftarrow{C}_\lambda^{trans}$ converges to 1 as these two measures are identical at the shortest-path extreme (the symmetry between $\vec{C}_\lambda^{trans}$ and $\overleftarrow{C}_\lambda^{trans}$ at the shortest path extreme will hold for any undirected network).

A node's propensity to be a costly transmission/informational source or target is projected onto the cortical surface in Fig 4, where we show the difference between a node's source and target costs for $\lambda_1=e^{-4.49}$, $\lambda_2=e^{-1.64}$, $\lambda_3=e^{0.37}$ and $\lambda_4=,e^{1.79}$ (same values highlighted in Fig 2 and Fig 3) . Cortical regions that are costly sources (compared to the cost of being a target) are colored red whereas regions that are costly targets (compared to the cost of being a source) are colored blue. To assess whether a region's propensity to be a costlier source or target is significant or not, given the node degrees and strengths, we standardize the empirical cost differences (i.e. $\vec{C}_\lambda^{trans}$-$\overleftarrow{C}_\lambda^{trans}$ and $\vec{C}_\lambda^{info}$-$\overleftarrow{C}_\lambda^{info}$) with respect to the distribution of cost differences computed on the ensembles of randomized networks, and test whether the empirically measured source and target cost difference is significantly larger than the difference measured in the randomized ensembles. Our results indicate that the right and left hemisphere superior parietal regions, the precuneus, and the fusiform gyri are significantly costlier sources, in terms of transmission cost, for $\lambda > e^{-0.8}$. Regions such as the right insula and rostral middle frontal cortex, right and left superior frontal cortex, and the precentral gyri are significantly costlier targets in terms of transmission cost, for $\lambda > e^{-0.8}$. In terms of informational cost, we find that the precentral gyri, paracentral lobule, right lateral-occipital cortex and left lingual gyrus are costlier sources, whereas the right posterior cingulate and supramarginal gyrus, left precuneus, right and left frontal pole, superior parietal cortex, and inferior parietal lobules are significantly costlier targets, for $\lambda < e^{-1.1}$. All z-scored cost differences as a function of λ are shown in Supplementary Fig S5 and S6.

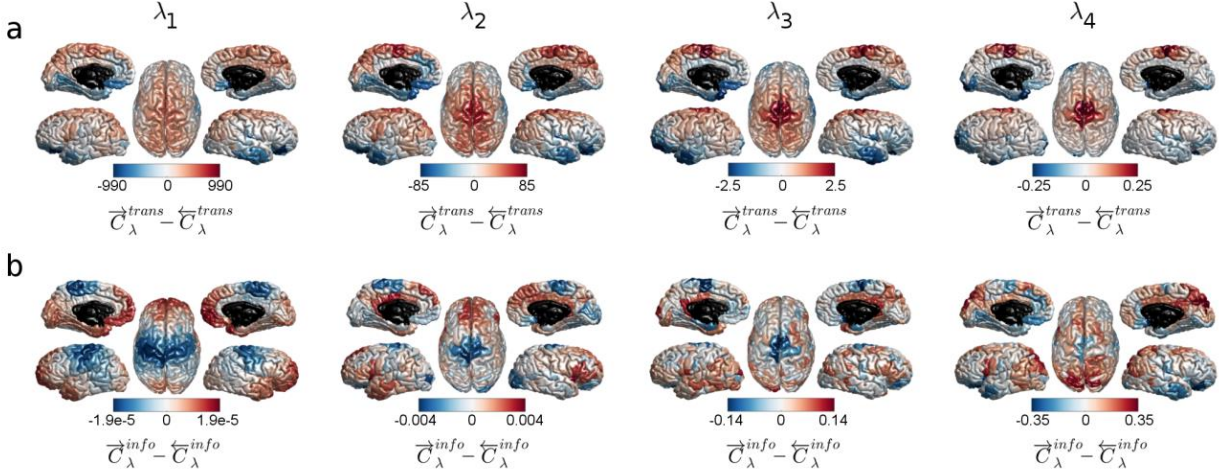

Fig 4. A brain region's propensity to be a costly source or target. Cortical surfaces show the difference between a node's source and target transmission costs. (a) $\vec{C}_\lambda^{trans} - \overleftarrow{C}_\lambda^{trans}$ for routing strategies generated with the values $\lambda_1$, $\lambda_2$, $\lambda_3$ and $\lambda_4$. (b) $\vec{C}_\lambda^{info} - \overleftarrow{C}_\lambda^{info}$ for routing strategies generated with the values $\lambda_1$, $\lambda_2$, $\lambda_3$ and $\lambda_4$. Red colored areas on the cortical surfaces correspond to nodes whose source transmission/informational cost is higher than their target transmission/informational cost. Blue colored areas correspond to nodes whose target transmission/informational cost is higher than their source transmission/informational cost. In all panels, $\lambda_1 = e^{-4.49}$, $\lambda_2 = e^{-1.64}$, $\lambda_3 = e^{0.37}$ and $\lambda_4 = e^{1.79}$.

**Routing strategies for privileged nodes**

In this section we will explore a different scenario where, in the interest of economizing on informational cost, we allow only a subset of *privileged nodes* to have access to global information. We consider increasingly larger size sets of *r* privileged nodes that are able to reshape their routing strategies according to the influence of global information. Privileged nodes are selected according to different node centrality rankings. Given a centrality-based ranking of nodes, we generate routing strategies for the *r*-highest ranked (privileged) nodes according to the stochastic model, where λ is an attribute that only applies to the set of privileged nodes; all *non- privileged nodes'* routing strategies remain unbiased and are equal to $P^{ref}(\mathbf{X})$. The left and middle panel of Fig 5 show network average values of $C_\lambda^{trans}$ and $C_\lambda^{info}$ (median across all subjects) as a function of λ for varying fractions of *privileged nodes* that are selected according to various centrality-based rankings. The black dotted lines show $C_\lambda^{trans}$ and $C_\lambda^{info}$, respectively, for the case in which all nodes' routing strategies are biased.

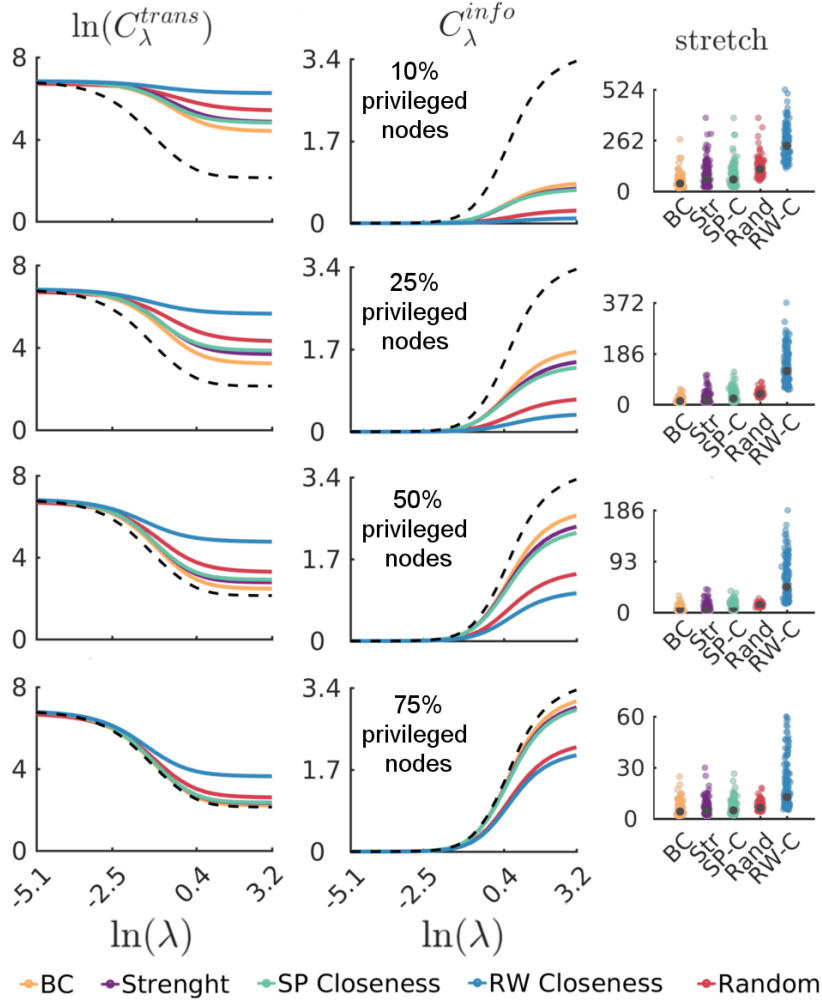

Fig 5. Network average values of $C_\lambda^{trans}$ (left panel) and $C_\lambda^{info}$ (middle panel) as a function of λ (node medians across all subjects) for 22, 55, 110, and 165 *privileged nodes* (corresponding to 10%, 25%, 50% and 75% of the network's nodes) that are selected according to betweenness centrality ranking (yellow line), strength ranking (purple line), shortest-path-based closeness centrality (green line), and random-walk-based closeness centrality (blue line). For comparison purposes, we also show cost measures for randomly sampled nodes (red line represents average across 500 samples). The dotted lines show $C_\lambda^{trans}$ and $C_\lambda^{info}$, respectively, for the case in which all nodes' routing strategies are biased (i.e. 100% privileged nodes). Right panel shows node stretch distributions for the different sets of privileged nodes and centrality rankings. Black markers indicate the median of the distributions.

This approach reveals three interesting properties about the routing capacity of the brain. First, the composition of the set of privileged nodes matters, as evidenced by the differences in $C_\lambda^{trans}$ and $C_\lambda^{info}$ that are obtained as the set size and composition is varied. Second, for a fixed number of privileged nodes, the more the system economizes on informational cost, the more it expends on transmission cost.

For example, routing strategies where we select privileged nodes according to betweenness centrality ranking yield smaller $C_\lambda^{trans}$ and larger $C_\lambda^{info}$ throughout the entire spectrum, compared to other centrality-based privileged node selections. Conversely, routing strategies where we select privileged nodes according to a random walk centrality ranking are the most costly in terms of $C_\lambda^{trans}$, but least costly in terms of $C_\lambda^{info}$. Third, a small number of privileged nodes can achieve a $C_\lambda^{trans}$ that is nearly as small as the $C_\lambda^{trans}$ achieved for shortest paths, but at a significantly smaller informational cost compared to what is needed to route messages through shortest-paths. Following [24] we define the stretch of a walk as the difference between the walk length and the shortest path length (both measured in terms of the number of edges/steps). Then, the stretch of a source node is the average stretch with respect to all target nodes in the network. For all sets of privileged nodes, we compute the stretch between the walks generated with $\lambda=e^{3.2}$ (i.e. the walks with the minimum $C_\lambda^{trans}$) and the shortest paths. Node stretch distributions (medians across all subjects) are shown in the right-side panel of Fig 5. For 25% privileged nodes, the median stretch for the BC ranking is 5.8; for the strength ranking it is 7; for the SP closeness ranking it is 7.1; for the random ranking it is 13.5; for the RW closeness it is 46.5. For 50% privileged nodes, the median stretch for the BC ranking is 4.2; for the strength ranking it is 4.7; for the SP closeness ranking it is 4.9; for the random ranking it is 6.4; for the RW closeness it is 12.7. Overall, these results indicate that efficient routing patterns can emerge even when less than half of the nodes are capable of routing information.

**A communication cost trade-off within subjects**

Our approach allows us to study the variability of communication cost measures across subjects. We first examine whether subjects who exhibit higher values of $C_\lambda^{trans}$ at $\lambda = 0$ (that is, longer walk lengths for the unbiased random walk) will maintain a high $C_\lambda^{trans}$ throughout the entire spectrum. Fig 6a shows correlations between all subject's $C_\lambda^{trans}$ across all values of λ. These correlations show that subjects who exhibit higher values of $C_\lambda^{trans}$ at $\lambda < e^{-3.1}$ are also subjects with the highest $C_\lambda^{trans}$ at λ >1, but the relationship is inverted in the middle of the spectrum.

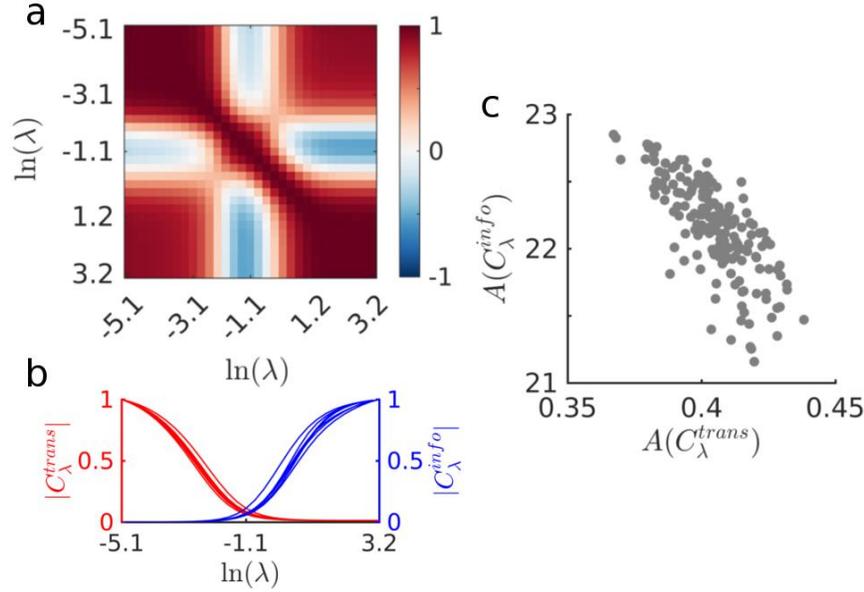

Fig 6. Communication cost trade-off within subjects. (a) Correlations between all subject's $C_\lambda^{trans}$ across all values of λ. Positive correlations are colored in red, negative correlations are colored in blue. (b) Eight subject's $C_\lambda^{trans}$ and $C_\lambda^{info}$ curves after normalization with respect to the max($C_\lambda^{trans}$) and max($C_\lambda^{info}$), respectively. Notice how some subject's $C_\lambda^{trans}$ curves decay faster than others, and how some subject's $C_\lambda^{info}$ curves grow faster than others. (c) Scatter plot of the computed areas under the normalized $C_\lambda^{trans}$ and $C_\lambda^{info}$ curves, sowing a trade-off between the decay of $C_\lambda^{trans}$ and the growth of $C_\lambda^{info}$ (the correlation between A($C_\lambda^{trans}$) and A($C_\lambda^{info}$) is $r = -0.74, p < 0.001$).

Finally, we investigate if there are differences in how individual subject's brain networks take advantage of the global information bias. We address this question by measuring the area under each subject's $C_\lambda^{trans}$ curve and $C_\lambda^{info}$ curve. Moreover, since we are interested in capturing the rate of decay and growth of subject's $C_\lambda^{trans}$ and $C_\lambda^{info}$ curves, we first normalize each subject's $C_\lambda^{trans}$ curve with respect to $C_\lambda^{trans}$ at λ = 0 (that is, the average length of unbiased random walks), and we normalize each subject's $C_\lambda^{info}$ curve with respect to $C_\lambda^{info}$ at λ = $e^{3.2}$ (that is, the max value of $C_\lambda^{info}$). The normalized $C_\lambda^{trans}$ and $C_\lambda^{info}$ curves of 8 subjects are shown in Fig 6b, illustrating curves that decay/grow faster with λ, which we can capture by measuring the area under the curve. Fig 6c shows a scatter plot of the areas under the normalized $C_\lambda^{trans}$ and $C_\lambda^{info}$ curves of all subjects, exhibiting a strong negative correlation between the normalized areas under $C_\lambda^{trans}$ and $C_\lambda^{info}$ ($r = -0.74$, $p<0.001$). This strong relationship indicates that there is a trade-off between a brain network's ability to take advantage of global information to route messages in a fast manner, and the amount of informational cost required to achieve optimally fast routing. How this trade-off is negotiated varies across individual subjects.

**Discussion**

The efficiency of communication in real world networks is not only determined by the speed with which messages are relayed, but the informational cost associated with selecting efficient routes is equally important. Here we introduce a stochastic model that generates routing strategies on a network by controlling the effect of global information over the actions of random walkers. We characterize the trade-offs between the cost of reshaping the system's dynamics ($C_\lambda^{info}$) and the cost of relaying messages through the network ($C_\lambda^{trans}$), and characterize these costs at a global, nodal and subject-wise level. Our results show that biased random walks can rapidly approach a shortest-path communication regime when afforded gradual small increases in the bias on global information.

Several properties inherent in this framework have important implications for the study of communication processes in brain networks. First, routing patterns are derived from a dynamical point of view, and not from a purely topological analysis of the system, allowing us to make use of well-established theoretical results about linear processes and biased random walks [25-27]. Second, routing patterns generated by these processes take place through multiple paths, combining structural robustness with tolerance to noisy interference. Third, routing strategies at each node are dynamic and allow us to model systems whose function demands flexible routing patterns that switch depending on intrinsic (e.g. amount of global information available) and/or extrinsic factors. Finally, building on the concept of dynamic routing patters, the notion of dynamic measures of centrality emerge naturally as a means to quantify the varying importance of nodes and edges under different underlying dynamics [30]. Here we have proposed the nodal cost measures $\vec{C}_\lambda^{trans}$, $\cev{C}_\lambda^{trans}$, $\vec{C}_\lambda^{info}$ and $\cev{C}_\lambda^{info}$ as dynamic source and target closeness centrality measures, but we note that additional centrality measures can be evaluated over the underlying *flow graphs* [26], that is, the weighted networks where the patterns of flow generated by our stochastic model are embedded.

The concept of communication dynamics has become increasingly important in the context of brain networks [40,41]. Here, we address some of the assumptions behind two widely used brain communication models, namely routing and diffusion models. On the one side, communication that takes place through shortest paths assumes that neural elements are able to identify the optimal path and route a signal/message through such path; however, the mechanisms by which signals are routed and the informational cost associated with routing them are rarely discussed. On the other side, communication that takes place through (unbiased) random walks assumes that signals are able to "bounce between nodes" for long periods of time. Yet, such a scheme raises issues about signal integrity and strength as well as metabolic cost. In our analysis, communication cost is not measured as a structural property of the network [17,23,36]. While wiring cost affects brain communication by means of being an important driver of brain geometry and network topology [1,17,15], it should be noted that wiring cost is a static property of the network (within relatively short time-scales) that is invariant under any communication process taking place on the network. In contrast, our framework approaches communication cost by considering two different cost components that are measured from the dynamics generated by a specific communication model. First, we consider the transmission cost which we interpret as a proxy for the metabolic cost of transmitting neural signals from one neural node to another. It has been estimated that about 50% of the brain's energy is used to drive signals across axons and synapses [1], suggesting that energy consumption is a strong incentive to minimize the length of communication pathways in neural systems. A natural derivation from the transmission cost measure results from its reciprocal (or inverse), thus extending and generalizing the global (or routing) efficiency [37] and diffusion efficiency [12,36] measures for shortest path and diffusion-based communication, respectively. Second, we consider the cost of reshaping the patterns of information flow (informational cost) that allow a signal to be efficiently routed towards a specific brain region. We conceptualize this cost as associated with modulatory processes that take place at the mesoscale or microscale, where signal traffic may be regulated as two neuronal population's firing rates change in order to synchronize and thus communicate [42], or as a process that emerges on top of the collective oscillatory dynamics of neural elements [43]. Hence, the communication cost measures proposed here are dynamic as they vary with the patterns of information flow generated by the communication process taking place on the network. Moreover, these cost measures intrinsically capture the informational cost associated with traversing high-degree nodes, that is, those comprising the brain's rich club. Indeed, it has been proposed that rich-club nodes facilitate integration of information within the network at the expense of a high wiring cost [23]; nonetheless, hubs are only advantageous for communication if signals can be routed through them, which implies high information cost [44]. Interestingly, a strong relationship between node degree, and the directionality with which signals are preferentially transferred through the network structure has been found in analytical, computational and empirical studies [45, 46], where it has been noted that high degree nodes' oscillatory activity lags in phase whereas low degree nodes' activity leads. These findings suggest that hub nodes tend to be directional targets, while low degree nodes act like sources [46]. Here we find similar routing patterns at the low-information end of the spectrum, where hub nodes tend to be low cost targets, while low degree nodes are efficient as sources. Building on these findings, our results also suggest that there is a regime within the spectrum where empirical networks are more efficient than their randomized counterparts; within this regime, the frontal cortex has an overabundance of high informational and high transmission cost source nodes; conversely, the parietal cortex exhibits an overabundance of high informational and high transmission cost target nodes.

Our results also contrast with well-established notions about the efficiency of random topologies [37,38], demonstrating that the randomized counterparts of empirical brain networks are only more efficient at the extremes of the communication spectrum. These findings have implications in terms of the use of randomized networks as a point of reference to normalize graph-efficiency measures [12,47]. Furthermore, our findings support the idea that the brain's topology does not only optimize a trade-off between wiring cost and efficient communication [15], but informational cost, the ability to access multiple pathways, and flexible routing patterns are additional important factors driving the network's organization.

Our findings regarding the selection of privileged nodes that have access to global information show that some nodes are poised to take advantage of global information more efficiently than others; in brain networks, efficient routing patterns can be achieved by allowing as few as 25% of the highest betweenness or strength centrality nodes to reshape their routing strategies according to a bias on global information. These results offer a new perspective on the role of highly central nodes in facilitating the co-existence of functional integration and segregation between and within neural sub-systems: densely connected clusters of nodes (network communities) tend to "trap" random walkers [48] which promotes their segregation, while a few well-connected privileged nodes are specialized to direct the sharing and exchange of information between clusters. Hence, the *privileged nodes* framework presented here may provide some insight about the underlying communication processes allowing the exchange of information between sub-systems [49,50].

Some limitations are worth mentioning. First, for this study, our application of the stochastic model is limited by restricting λ to be a global attribute for all nodes, or for a set of privileged nodes; nonetheless, it is feasible (although computationally intensive) and perhaps more realistic to define λ as a continuously varying nodal property, λ(*i*). Second, the stochastic model considers a scenario where communication between all nodes and a given target is equally salient. In systems such as the brain, where different sub-systems are associated with specific cognitive tasks, it is unlikely that all node pairs require the ability to efficiently exchange information with all other nodes. In this sense, the cost measures computed here may serve as an upper bound for the actual communication cost. Third, linear dynamics may not be appropriate for systems that exhibit highly complex non-linear dynamics. Indeed, the brain is highly complex, topologically and dynamically. Yet, its complexity allows us to study it at different scales [51]. While it is clear that both structure and dynamics must be considered simultaneously to achieve a more comprehensive description of the system, it is still unclear how communication dynamics manifest at the various scales at which we are able to capture brain structure and dynamics. Hence, there is no evidence to discard linear dynamics as good approximation of the routing patterns taking place on large-scale brain networks.

Taken together, our work establishes a theoretical framework to study the efficiency of a broad range of communication processes on complex networks. While we have focused on a particular class of biased random walks where biases depend on the topological distance to target nodes, we note that biases may also depend on other aspects of the global topology or the embedding of a network in physical space [14,28]. Overall, this framework can be used to study any real world network that employs communication or navigation processes in its operation. It may be used, for instance, to infer pathways through which information is preferentially transferred, or, when such pathways are known, to infer the search and navigation strategies that allow accessing these pathways. In the context of brain networks, this theoretical framework may prove useful to identify efficient communication strategies that balance different aspects of the cost associated with neural communication.

**Materials and Methods**

**Data sets.**

**LAU.** Informed written consent in accordance with the Institutional guidelines (protocol approved by the Ethics Committee of Clinical Research of the Faculty of Biology and Medicine, University of Lausanne, Switzerland) was obtained for all subjects. Forty healthy subjects (16 females; 25.3 ± 4.9 years old) underwent an MRI session on a 3T Siemens Trio scanner with a 32-channel head coil. Magnetization prepared rapid acquisition with gradient echo (MPRAGE) sequence was 1-mm in-plane resolution and 1.2-mm slice thickness. DSI sequence included 128 diffusion weighted volumes + 1 reference b0 volume, maximum *b* value 8000 s/mm$^2$, and 2.2 × 2.2 × 3.0 mm voxel size. EPI sequence was 3.3-mm in-plane resolution and 3.3-mm slice thickness with TR 1920 ms. DSI and MPRAGE data were processed using the Connectome Mapper Toolkit [52]. Each participant's gray and white matter compartments were segmented from the MPRAGE volume. The grey matter volume was subdivided into 68 cortical and 15 subcortical anatomical regions, according to the Desikan-Killiany atlas, defining 83 anatomical regions. These regions were hierarchically subdivided to obtain five parcellations, corresponding to five different scales [53]. The present study uses a parcellation comprising 233 regions of interest (ROI). Whole brain deterministic streamline tractography was performed on reconstructed DSI data, initiating 32 streamline propagations (seeds) per diffusion direction, per white matter voxel [54]. Within each voxel, seeds were randomly placed and for each seed, a fiber streamline was grown in two opposite directions with a 1mm fixed step. Fibers were stopped if a change in direction was greater than 60 degrees/mm. The process was complete when both ends of the fiber left the white matter mask. For each individual subject, connection weights between pairs of ROI are quantified as a fiber density [55]. Thus, the connection weight between the pair of brain regions $\{u,v\}$ captures the average number of streamlines per unit surface between *u* and *v*, corrected by the average length of the streamlines connecting such brain regions. The aim of these corrections is to control for the variability in cortical region size and the linear bias toward longer streamlines introduced by the tractography algorithm. Fiber densities were used to construct individual subject structural connectivity matrices. Each structural connectivity matrix is then modeled as the adjacency matrix $A=\{a_{ij}\}$ of a graph $G = \{V,G\}$ with nodes $V = \{v_1,...,v_n\}$ representing ROIs, and weighted, undirected edges $E = \{e_1,...,e_m\}$ representing anatomical connections with their fiber densities.

**HCP.** High-resolution diffusion-weighted (DWI) data from the Human Connectome Project [34] including 173 subjects (Q3 release; males and females mixed, age 22–35 years; imaging parameters: voxel size 1.25 mm isotropic, TR/TE 5520/89.5 ms, 90 diffusion directions with diffusion weighting 1000, 2000, or 3000 s/mm$^2$) was used to reconstruct macroscale human connectomes for each subject. DWI data processing included the following: (1) eddy current and susceptibility distortion correction, (2) reconstruction of the voxelwise diffusion profile using generalized q-sampling imaging, and (3) whole-brain streamline tractography (see ref 56 for details).

Cortical segmentation and parcellation was performed on the basis of a high-resolution T1-weighted image (voxel size: 0.7 mm isotropic) using FreeSurfer [57], automatically parcellating the complete cortical sheet into 219 distinct regions using a subdivision of the Desikan-Killiany atlas. White matter pathways were reconstructed using generalized Q-sampling imaging (GQI), and streamline tractography [56]. A streamline was started in each white matter voxel, following the most matching diffusion direction from voxel to voxel until a streamline reached the gray matter, exited the brain tissue, made a turn of >45 degrees or reached a voxel with a low fractional anisotropy (<0.1). For each individual subject, a 219 x 219 weighted connectivity matrix was constructed by taking the strength of reconstructed region-to-region connections as the number of tractography streamlines between i and j, and dividing by the average cortical surface area of both regions [53].

**Defining topological distances for human structural connectivity networks.** The edge weights of human brain structural connectivity networks are normally defined in terms of proximity measures such as the number of streamlines or fiber densities. These proximity edge-weights are often interpreted as a measure of information flow or traffic capacity that can travel through a connection (a notion that is analogous to the concept of bandwidth in telecommunication networks). Hence, the proximity between two brain regions is determined by the sequence of edges that maximize the traffic or flow capacity. In order to define topological distances on human brain structural connectivity networks, a proximity-to-distance mapping must be applied over the set of edge-weights, such that large edge-weights (large edge-proximities) are mapped onto small edge-distances, and small edge-weights are mapped onto large edge-distances. The proximity-to-distance mapping can be defined in various ways. Following previous work [6,44], in this study we use the mapping $d_{ij} = log(1/w_{ij})$, where $w_{ij}$ are edge-proximities (i.e. fiber densities) and $d_{ij}$ are the resulting edge-distances. This mapping yields edge-distances with a log-normal distribution, which is consistent with evidence showing log-normal distributions of synaptic strengths between cortical cells [58] and cortico-cortical projections [59]. Finally, in order to implement this mapping, we first normalize all edge-weights, to ensure that $w_{ij}$ are bounded in the interval [0,1]. As shown previously [32], there is a unique linear function that can normalize any weighted graph onto the unit interval without affecting network properties:

$$\overline{w_{ij}} = \frac{(1 - 2\epsilon)w_{ij} + (2\epsilon - 1) \cdot MIN(w_{ij})}{MAX(w_{ij}) - MIN(w_{ij})} + \epsilon$$

Here we use $\epsilon=MIN(w_{ij})$, in order to obtain normalized edge-weights in the interval (0,1) which allows us to apply the proximity-to-distance map $d_{ij} = log(1/\overline{w_{ij}})$.

**Computation of $n_\lambda^t$.** Let $\mathbf{M} = \{\mathbf{S}, \mathbf{P}_\lambda\}$ be a Markov chain composed by a set of $N$ states $\mathbf{S}=\{1,2,..., i_N\}$ that correspond element by element to the set of nodes of a graph $G$ with $N$ nodes and $E$ edges; $\mathbf{P}_\lambda$ is the matrix of transition probabilities characterizing the probability of transitioning from one state to another. Then, $\mathbf{P}_\lambda(i,j) \neq 0$ if and only if an edge exists between nodes $i$ and $j$ in graph $G$.

Let $\mathbf{X}$ be a random variable indicating the current state of the chain, or equivalently, the current node where the walker is located; $\mathbf{Y}$ is the random variable indicating the node to which the walker will move in the next time step, and $\mathbf{T}$ is the random variable indicating the target node where the walk will terminate (we assume that $\mathbf{M}$ is an irreducible chain). For a given value of $\lambda$, and an specified target $\mathbf{T} = t$, let $\mathbf{P}_\lambda$ be the $N$x$N$ matrix of transition probabilities where elements of $\mathbf{P}_\lambda$ are defined as

$$P_\lambda(Y = j \mid X = i, T = t) = \exp(-(\lambda(d_{ij} + g_{jt}) + d_{ij}))\frac{1}{Z_i^t}$$

where $Z_i^t = \sum_j \exp(-(\lambda(d_{ij} + g_{jt}) + d_{ij}))$ is a normalization factor, $d_{ij}$ is the distance from $i$ to $j$ and $g_{jt}$ is the geodesic distance from $j$ to the target node $t$.

We make $\mathbf{M}$ an absorbing chain and $t$ an absorbing state by setting all transition probabilities $P_\lambda(\mathbf{Y}=j|\mathbf{X}=t,\mathbf{T}=t) = 0$ for $j \neq t$ and $P_\lambda(\mathbf{Y}=j|\mathbf{X}=t,\mathbf{T}=t) = 1$ for $j = t$, and define $\mathbf{Q}^t_\lambda$ as the $N$-1x$N$-1 matrix of transition probabilities from non-absorbing to non-absorbing states. Then, $\mathbf{n}^t_\lambda= (\mathbf{I}- \mathbf{Q}^t_\lambda)^{-1}$ is the fundamental matrix for the absorbing chain [31], and the elements $n^t_\lambda(i,j)$ denote the amount of time that the chain spends in the j-th non-absorbing state when the chain is initialized in the i-th non-absorbing state. In other words, if we take $\mathbf{P}_\lambda$ to represent the transition probabilities for a (biased) random walker on graph $G$, and going from a source node $i$ to a target node $t$, then $n^t_\lambda(i,j)$ represents the number of times that the random walker starting at node $i$ visits node $j$ before it reaches node $t$.

**Transition probabilities for degenerate paths**. Let $\pi_1$ and $\pi_2$ be any two paths going from node $i$ to node $t$ through edges $\{i,j\}$, and $\{i,k\}$, respectively. The ratio between the transition probabilities $P^t_{ij}$ and $P^t_{ik}$ is:

$$\frac{P^t_{ij}}{P^t_{ik}} = \frac{\exp\left(-\lambda(d_{ij} + g_{jt})\right)\exp(-d_{ij})}{\exp\left(-\lambda(d_{ik} + g_{kt})\right)\exp(-d_{ik})}$$

Assume that the length of $\pi_1$ and $\pi_2$ is equal, so $d_{ij}+g_{jt} = d_{ik}+g_{kt}$. Then we can write:

$$\frac{P^t_{ij}}{P^t_{ik}} = \frac{\exp(-d_{ij})}{\exp(-d_{ik})}$$

Now, let $S$ indicate the set of edges leaving from node $i$ along which there is a shortest path from node $i$ to node $t$. Since all edges in $S$ lie on shortest paths, for any pair of edges $\{i,j\}, \{i,k\} \in S$, it must be that $d_{ij}+g_{jt} = d_{ik}+g_{kt}$. Then, when $\lambda \to \infty$, we can write

$$P_{ij}^t = \begin{cases} \frac{\exp(-d_{ij})}{\sum_{\{i,j'\} \in S} \exp(-d_{ij'})} & if\ \{i,j\} \in S \\ 0 & otherwise \end{cases}$$

If the network is unweighted, then all $d_{ij}$ = const. In that case, all edges in $S$ will have a uniform transition probability from node $i$.

Note that in the $\lambda \to \infty$ case, only transitions along shortest paths will be allowed. This means that the random walk path lengths will be equal to shortest path lengths.

**Randomized Networks.** For each subject, we created a population of 500 randomized brain networks, with preserved degree strength sequence, and preserved weight distribution, following the procedure described in [60]. The empirical networks were first binarized and then randomized by swapping pairs of connections as described in [12], thus preserving the binary degree of each node. In order to approximate the strength sequence of the empirical structural connectivity matrices, we used a simulated annealing algorithm that minimizes the cost function $C = \sum_i |s_i - r_i|$, where $s_i$ is the strength of node $i$ in the empirical network and $r_i$ is the strength in the randomized network. The cost function is minimized by randomly permuting weight assignments across edges and probabilistically accepting the permutations that reduced the energy as the temperature parameter of the algorithm is decreased. The annealing schedule consisted of 123 iterations and a starting temperature of $t_0$=100, which was scaled by 0.125 after each iteration. The result of this procedure was an average final energy of $C$=0.2797±0.04, which indicates that the average strength discrepancy per node was between 0.0011 - 0.0014.

**Intrinsic Connectivity Networks.** We mapped the Desikan Killiany anatomical parcels used to construct individual subject structural connectivity networks, onto the seven intrinsic connectivity networks (ICN) defined by Yeo et al. (2011) [61]. This parcellation was derived by using a clustering algorithm to partition the cerebral cortex of 1000 healthy subjects into networks of functionally coupled regions. The clustering procedure resulted in the definition of seven clusters comprising systems previously described in the literature including the visual (VIS) and somatomotor (SM) regions, dorsal (DA) and ventral (VA) attention networks, frontoparietal control (FP), limbic (LIM) and default mode network (DMN). The mapping between the Desikan-Killiany anatomical parcels and the seven ICNs from the ICN parcellation was obtained by extracting the vertices of the brain surface corresponding to each anatomical region in the Desikan-Killiany atlas, and then evaluating the mode of the vertices' assignment in the ICN parcellation

Acknowledgments. The authors thank Alessandra Griffa for providing the processed DTI data as well as computational tools for visualization. We also thank Filip Miscevic and Bratislav Mišić for fruitful discussions. O.S. was supported by the National Institutes of Health (R01-AT009036). P.H. was supported by the Leenaards Foundation.

Conflict of Interest.
The authors have no competing interests.

Author Contributions. A.A-K and A.K. designed the experiments; A.A-K. performed the computational analysis; A.A-K, A.K, X.Y and O.S. Analyzed the results; M.van den H. and P.H. provided the data; A.A-K., X.Y., A.K. and O.S. Wrote and prepared the manuscript; all authors reviewed and edited the manuscript.

SUPPLEMENTAL INFORMATION

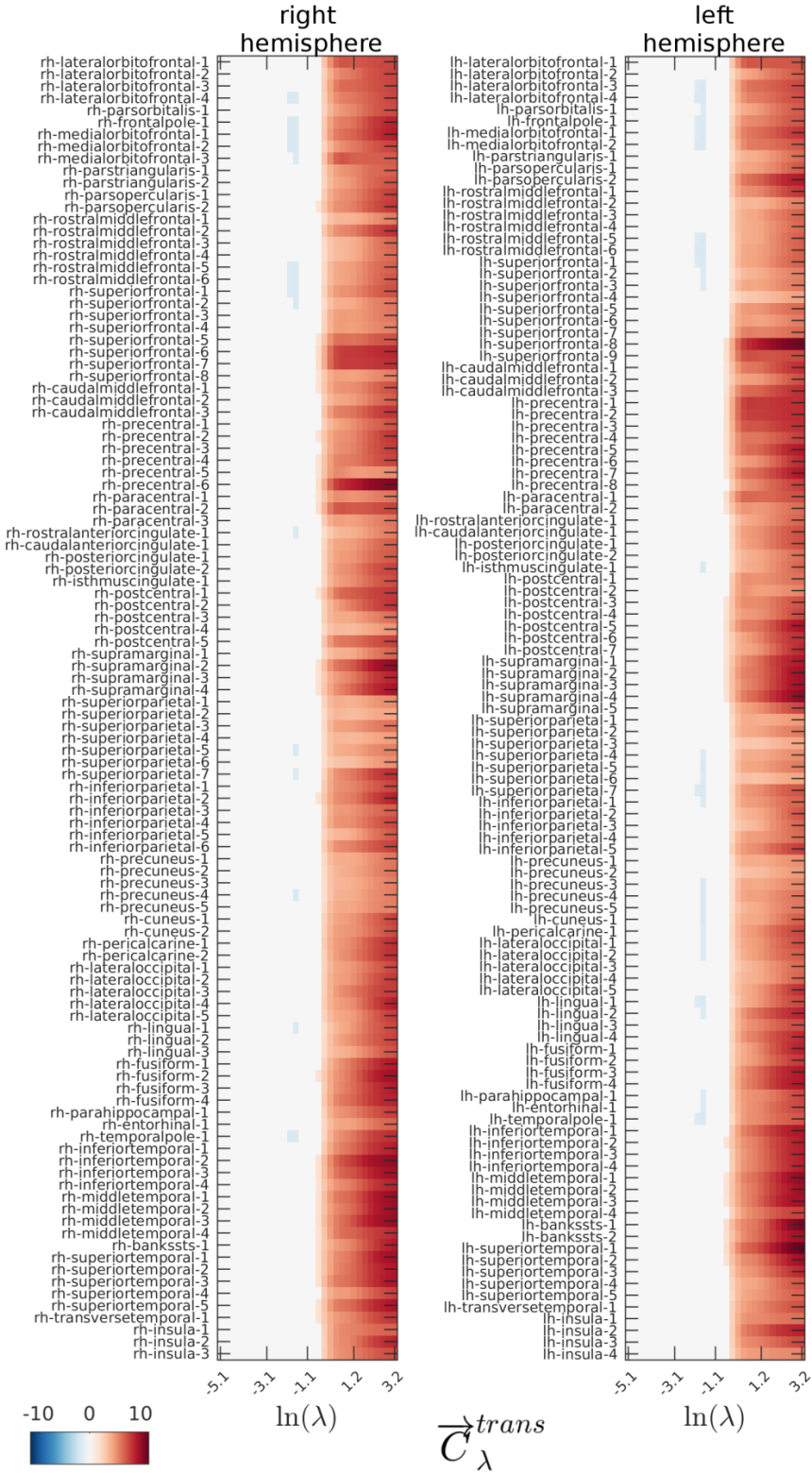

**Fig S1. z-scored source transmission costs as a function of** λ. For each subject's structural connectivity network, the source transmission cost ($\vec{C}_\lambda^{trans}$) of every node was standarized with respect to the corresponding distribution of $\vec{C}_\lambda^{trans}$ measured on an ensemble of 500 randomized networks. z-scored values were then thresholded according to a type I error *α = 0.01*. Red regions on the spectrum indicate nodes with significantly high $\vec{C}_\lambda^{trans}$, whereas blue regions on the spectrum indicate nodes with significantly low $\vec{C}_\lambda^{trans}$.

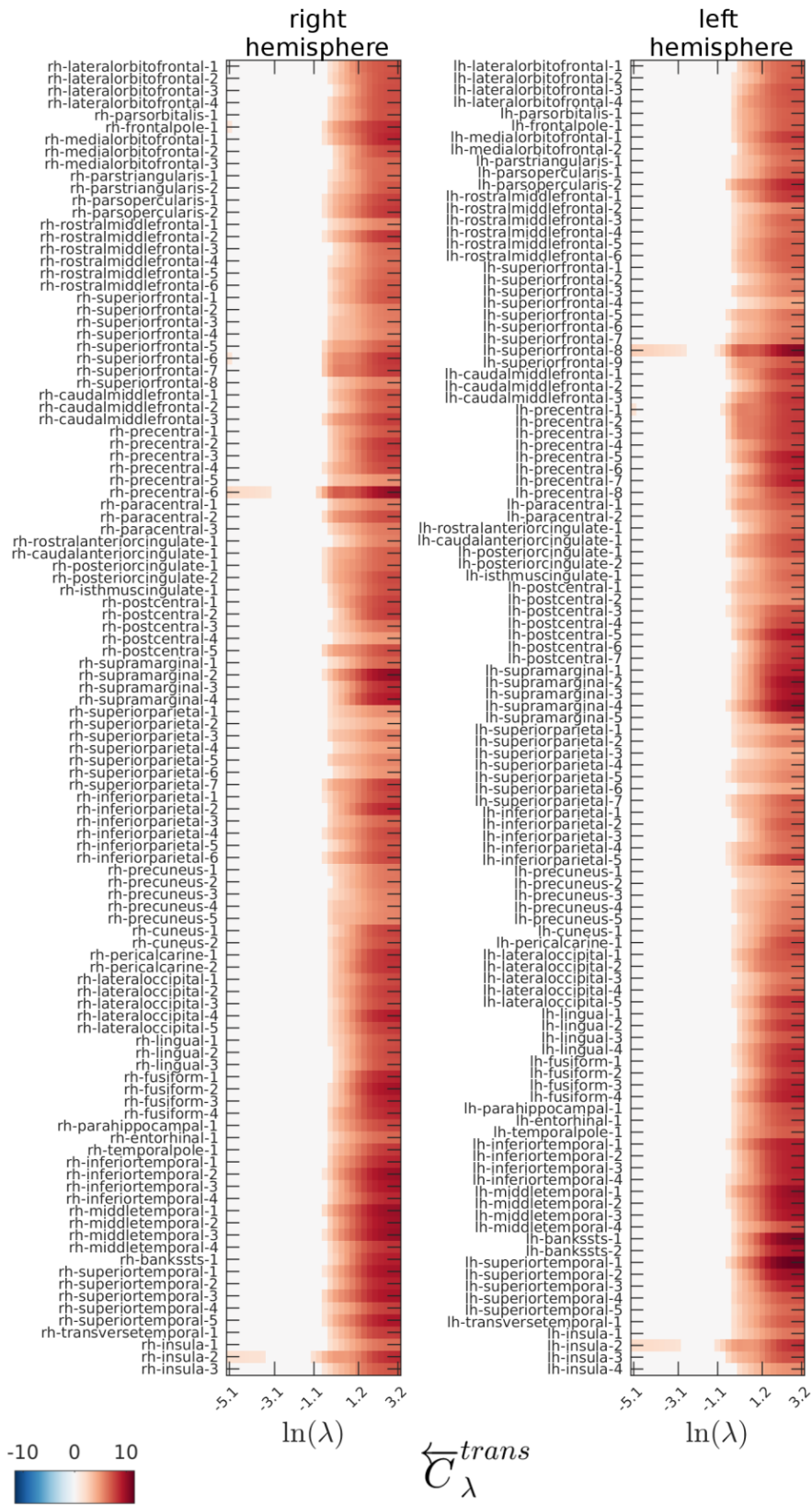

**Fig S2. z-scored target transmission costs as a function of λ.** For each subject's structural connectivity network, the target transmission cost ($\overleftarrow{C}_\lambda^{trans}$) of every node was standarized with respect to the corresponding distribution of $\overleftarrow{C}_\lambda^{trans}$ measured on an ensemble of 500 randomized networks. z-scored values were then thresholded according to a type I error $\alpha = 0.01$. Red regions on the spectrum indicate nodes with significantly high $\overleftarrow{C}_\lambda^{trans}$, whereas blue regions on the spectrum indicate nodes with significantly low $\overleftarrow{C}_\lambda^{trans}$.

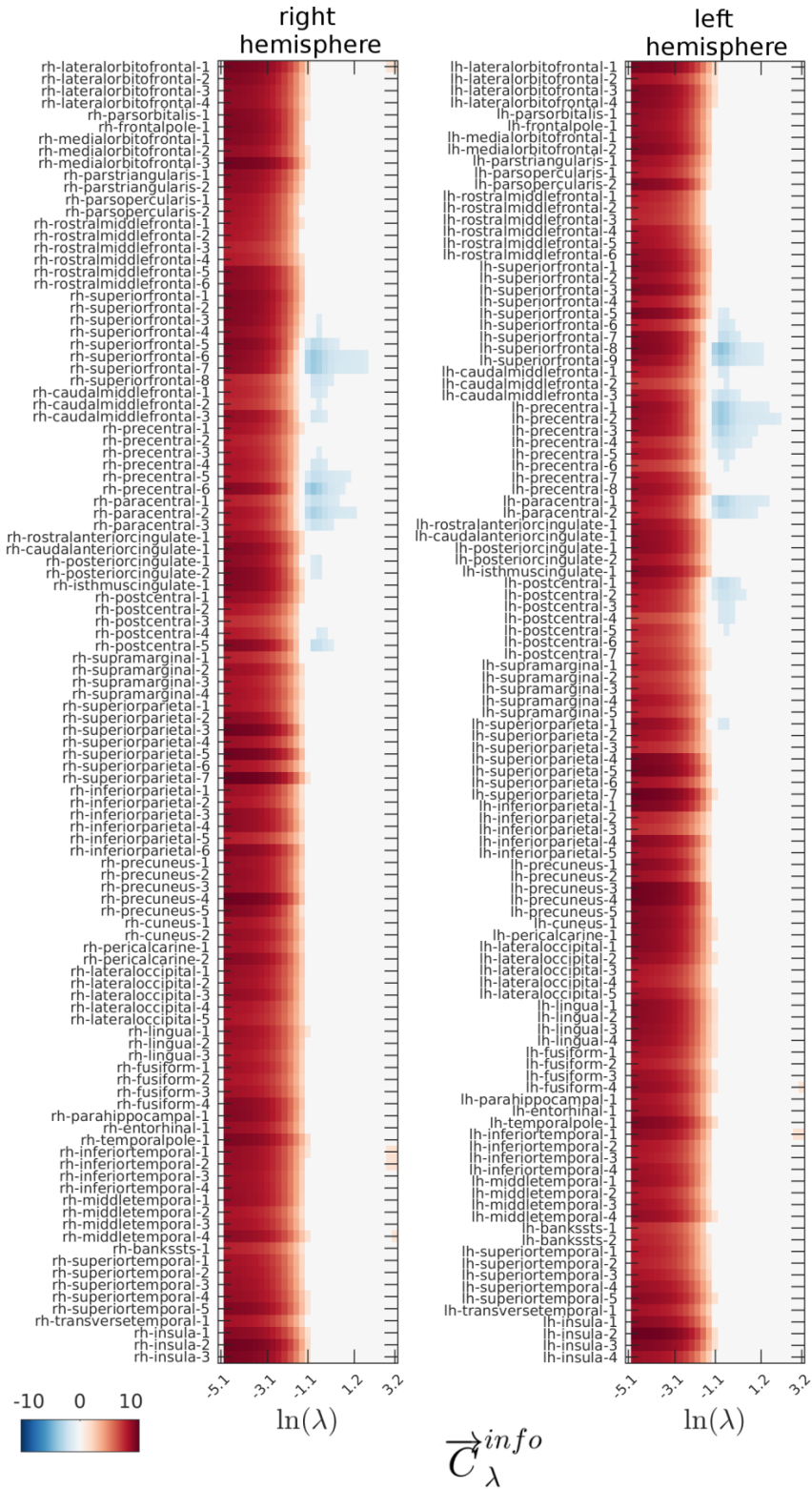

**Fig S3. z-scored source informational costs as a function of λ.** For each subject's structural connectivity network, the source informational cost ($\vec{C}_\lambda^{info}$) of every node was standardized with respect to the corresponding distribution of $\vec{C}_\lambda^{info}$ measured on an ensemble of 500 randomized networks. z-scored values were then thresholded according to a type I error *α = 0.01*. Red regions on the spectrum indicate nodes with significantly high $\vec{C}_\lambda^{info}$, whereas blue regions on the spectrum indicate nodes with significantly low $\vec{C}_\lambda^{info}$.

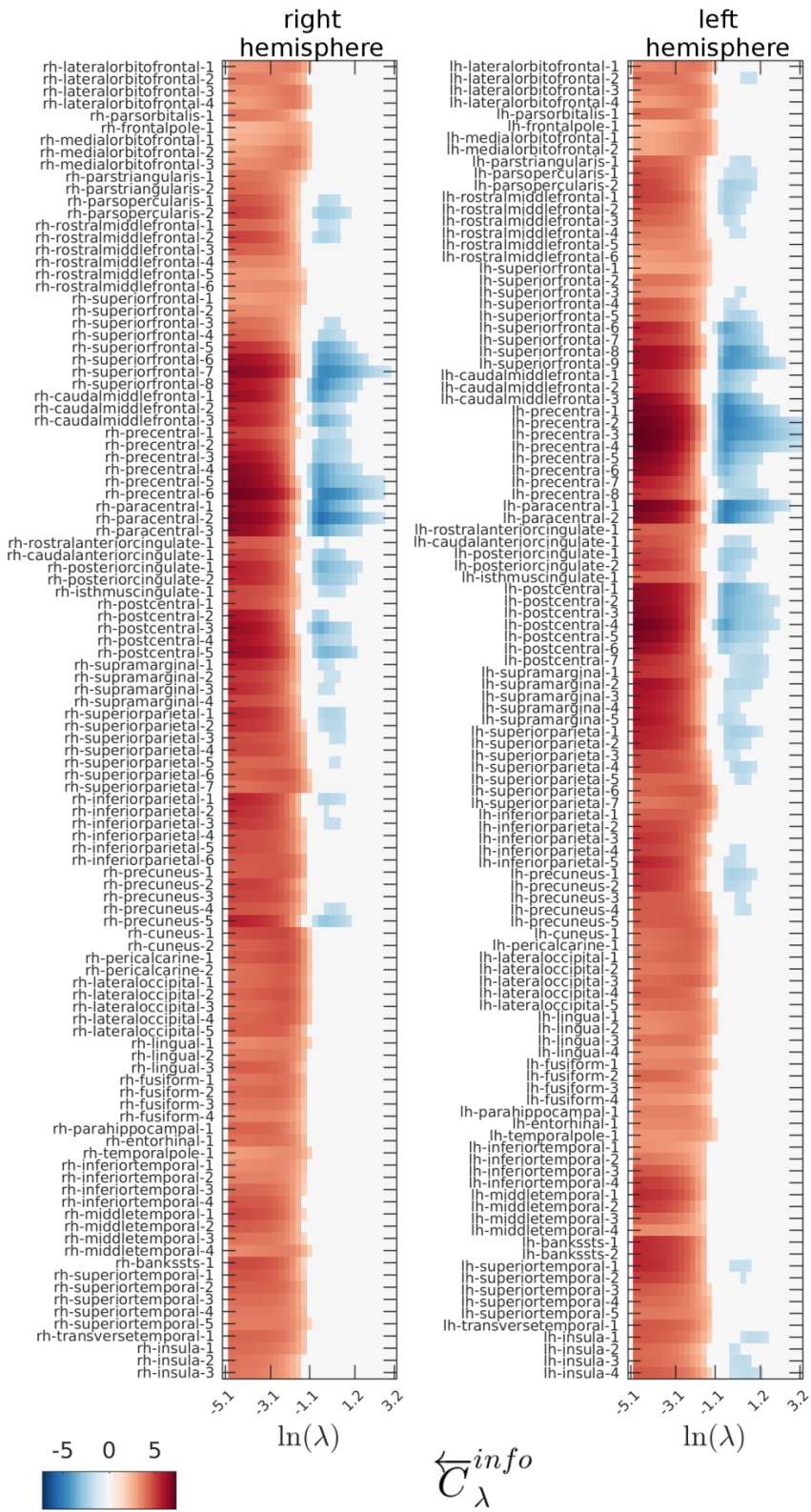

**Fig S4. z-scored target informational costs as a function of** λ**.** For each subject's structural connectivity network, the target informational cost ($\overleftarrow{C}_\lambda^{info}$) of every node was standarized with respect to the corresponding distribution of $\overleftarrow{C}_\lambda^{info}$ measured on an ensemble of 500 randomized networks. z-scored values were then thresholded according to a type I error $\alpha = 0.01$. Red regions on the spectrum indicate nodes with significantly high $\overleftarrow{C}_\lambda^{info}$, whereas blue regions on the spectrum indicate nodes with significantly low $\overleftarrow{C}_\lambda^{info}$.

$$\overrightarrow{C}_\lambda^{trans} - \overleftarrow{C}_\lambda^{trans}$$

**Fig S5. z-scored difference between source and target transmission costs as a function of λ.** For each subject's structural connectivity network, the difference between the source and target transmission cost ($\vec{C}_\lambda^{trans}$-$\cev{C}_\lambda^{trans}$) of every node was standardized with respect to the corresponding distribution of $\vec{C}_\lambda^{trans}$-$\cev{C}_\lambda^{trans}$ measured on an ensemble of 500 randomized networks. A one-tailed test (*α = 0.01*) was performed to test the hypothesis that the difference $\vec{C}_\lambda^{trans}$-$\cev{C}_\lambda^{trans}$> *null*($\vec{C}_\lambda^{trans}$-$\cev{C}_\lambda^{trans}$) if $\vec{C}_\lambda^{trans}$-$\cev{C}_\lambda^{trans}$>0, or $\vec{C}_\lambda^{trans}$-$\cev{C}_\lambda^{trans}$< *null* ($\vec{C}_\lambda^{trans}$-$\cev{C}_\lambda^{trans}$) if $\vec{C}_\lambda^{trans}$-$\cev{C}_\lambda^{trans}$<0, where *null*($\vec{C}_\lambda^{trans}$-$\cev{C}_\lambda^{trans}$) is the mean difference obtained from the ensemble of randomized networks. Red regions on the spectrum indicate nodes with significantly high (>0) $\vec{C}_\lambda^{trans}$-$\cev{C}_\lambda^{trans}$, whereas blue regions on the spectrum indicate nodes with significantly small (<0) $\vec{C}_\lambda^{trans}$-$\cev{C}_\lambda^{trans}$.

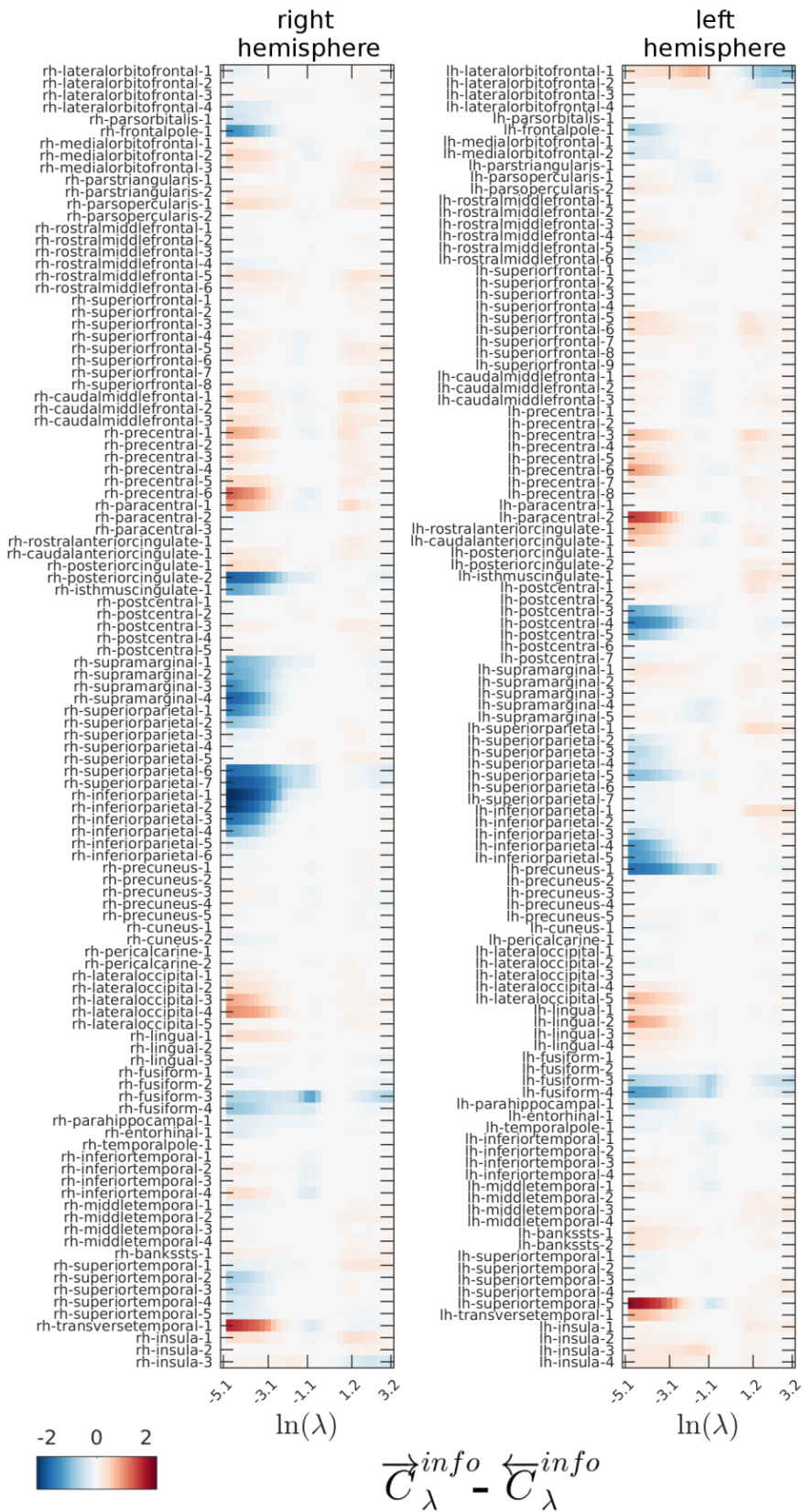

**Fig S6. z-scored difference between source and target informational costs as a function of λ.** For each subject's structural connectivity network, the difference between the source and target transmission cost ($\vec{C}_\lambda^{info}$-$\overleftarrow{C}_\lambda^{info}$) of every node was standardized with respect to the corresponding distribution of $\vec{C}_\lambda^{trans}$-$\overleftarrow{C}_\lambda^{trans}$ measured on an ensemble of 500 randomized networks. A one-tailed test ($\alpha = 0.01$) was performed to test the hypothesis that the difference $\vec{C}_\lambda^{info}$-$\overleftarrow{C}_\lambda^{info}$> $null(\vec{C}_\lambda^{info}$-$\overleftarrow{C}_\lambda^{info})$ if $\vec{C}_\lambda^{info}$-$\overleftarrow{C}_\lambda^{info}$>0, or $\vec{C}_\lambda^{info}$-$\overleftarrow{C}_\lambda^{info}$< $null(\vec{C}_\lambda^{info}$-$\overleftarrow{C}_\lambda^{info})$ if $\vec{C}_\lambda^{info}$-$\overleftarrow{C}_\lambda^{info}$<0, where $null(\vec{C}_\lambda^{info}$-$\overleftarrow{C}_\lambda^{info})$ is the mean difference obtained from the ensemble of randomized networks. Red regions on the spectrum indicate nodes with significantly high (>0) $\vec{C}_\lambda^{info}$-$\overleftarrow{C}_\lambda^{info}$, whereas blue regions on the spectrum indicate nodes with significantly small (<0) $\vec{C}_\lambda^{info}$-$\overleftarrow{C}_\lambda^{info}$.

Replication data set (LAU).

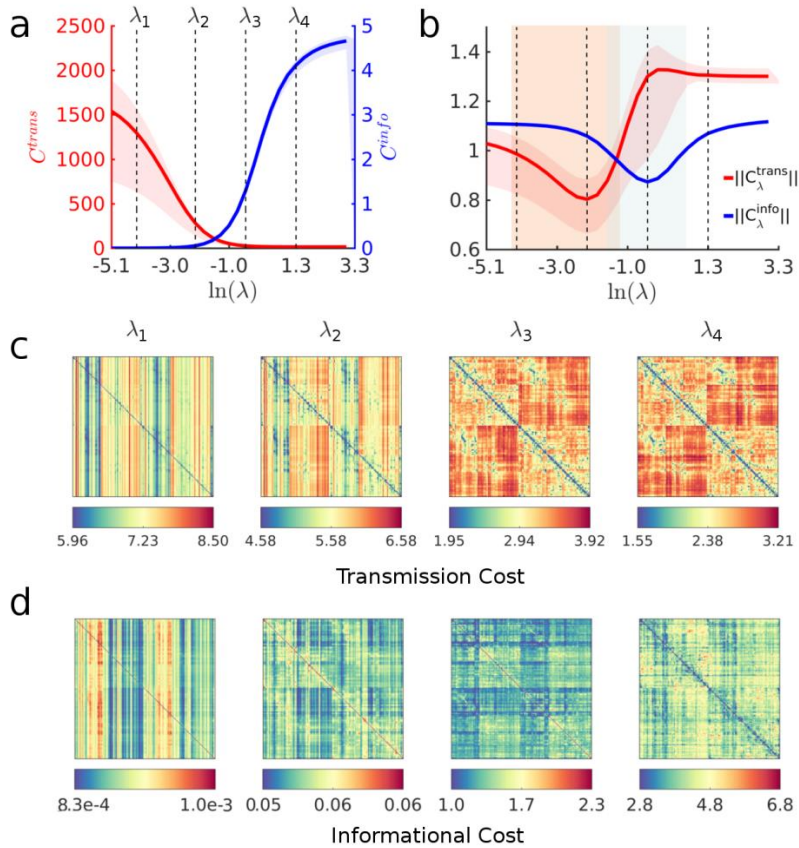

**Fig S7. A spectrum of communication processes.** (a) Averages of $C_\lambda^{trans}$ (red) and $C_\lambda^{info}$ (blue) across all node pairs, as a function of λ. Solid red and blue lines correspond to the median across all subjects, whereas the shaded red and blue regions denote the 95th percentile. Vertical dashed lines correspond to the values $\lambda_1=e^{-4.19}$, $\lambda_2=e^{-2.16}$, $\lambda_3=e^{-.042}$ and $\lambda_4=,e^{1.31}$. (b) Averages of $\|C_\lambda^{trans}\|$ (red) and $\|C_\lambda^{info}\|$ (blue) across all node pairs. These curves are computed by normalizing $C_\lambda^{trans}$ and $C_\lambda^{info}$ with respect to the same cost measures computed on ensembles of 500 randomized networks (per subject). Shaded red and blue areas indicate sections of the curves $\|C_\lambda^{trans}\|$ and $\|C_\lambda^{info}\|$ that are smaller than 1, respectively, indicating the regions in the spectrum where the communication cost of empirical networks is smaller than the cost computed on the randomized ensembles. The dashed vertical lines are placed at the minimum and maximum of $\|C_\lambda^{trans}\|$ ($\lambda_2$ and $\lambda_3$, respectively), and at two points near the extremes of the spectrum (($\lambda_1$ and $\lambda_4$). (c) pairwise values of $C_\lambda^{trans}(i,t)$ for all node pairs. (d) pairwise values of $C_\lambda^{info}(i,t)$ for all node pairs. For the panels $\lambda_1=e^{-4.19}$, $\lambda_2=e^{-2.16}$, $\lambda_3=e^{-.042}$ and $\lambda_4=e^{1.31}$.

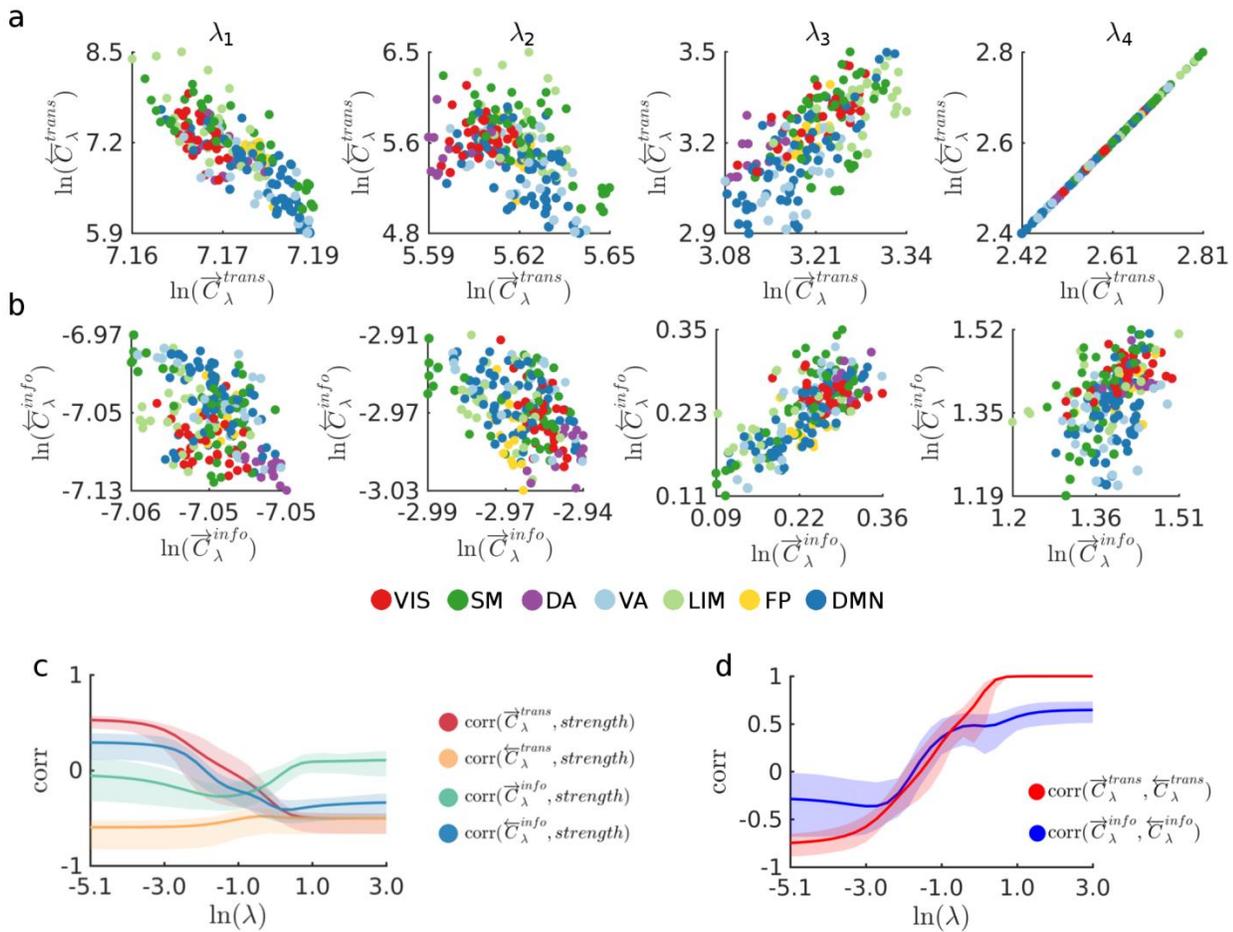

**Fig S8. Nodal average transmission costs for four increasingly biased routing strategies.** (a) Scatter plots show the transmission cost associated to each node when it acts as source ($\vec{C}_\lambda^{trans}$) and target ($\overleftarrow{C}_\lambda^{trans}$) during communication processes taking place under routing strategies generated with the values $\lambda_1=e^{-4.19}$, $\lambda_2=e^{-2.16}$, $\lambda_3=e^{-.042}$ and $\lambda_4=e^{1.31}$. (b) Scatter plots show the transmission cost associated to each node when it acts as source ($\vec{C}_\lambda^{info}$)

and target ($\overleftarrow{C}_\lambda^{info}$) during communication processes taking place under routing strategies generated with the values $\lambda_1$, $\lambda_2$, $\lambda_3$ and $\lambda_4$. Markers in the scatter plots in (a) and (b), representing each node, are colored according to the node's functional role according to the 7 intrinsic connectivity networks (ICN) defined by Yeo et al. (2011) (61): Visual (VIS), Somatomotor (SM), Dorsal Attention (DA), Ventral Attention (VA), Limbic (LIM), Frontal Parietal (FP), and Default Mode Network (DMN) (see Figure SI1 showing ICNs projected on a cortical surface). The size of the markers is proportional to node's strength. (c) Correlations between node strength and $\vec{C}_\lambda^{trans}$ (red), $\overleftarrow{C}_\lambda^{trans}$ (orange), $\vec{C}_\lambda^{info}$ (green) and $\overleftarrow{C}_\lambda^{info}$ (blue) as a function of λ. Solid lines show median correlation across all subjects, shaded areas surrounding the lines show 95$^{th}$ percentile. Shaded colored areas between the vertical dashed lines indicate regions where the correlations were not significant (p > 0.001). (d) Correlation between $\vec{C}_\lambda^{trans}$ and $\overleftarrow{C}_\lambda^{trans}$ (red), and $\vec{C}_\lambda^{info}$ and $\overleftarrow{C}_\lambda^{info}$ (blue), as a function of λ. Solid lines show medians across all subjects and shaded areas surrounding solid lines show the 95$^{th}$ percentile. Shaded areas between the vertical dashed lines indicate areas where correlation values were not significant (p > 0.001).

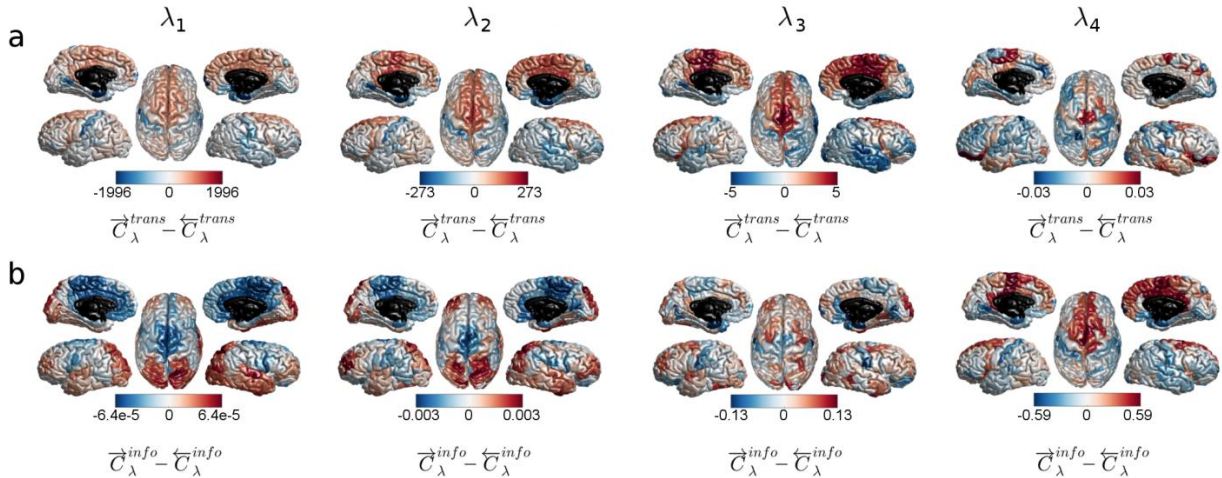

**Fig S9. A brain region's propensity to be a costly source or target.** Cortical surfaces show the difference between a node's source and target transmission costs. (a) $\vec{C}_\lambda^{trans} - \overleftarrow{C}_\lambda^{trans}$ for routing strategies generated with the values $\lambda_1$, $\lambda_2$, $\lambda_3$ and $\lambda_4$. (b) $\vec{C}_\lambda^{info} - \overleftarrow{C}_\lambda^{info}$ for routing strategies generated with the values $\lambda_1$, $\lambda_2$, $\lambda_3$ and $\lambda_4$. Red colored areas on the cortical surfaces correspond to nodes whose source transmission/informational cost is higher than their target transmission/informational cost. Blue colored areas correspond to nodes whose target transmission/informational cost is higher than their source transmission/informational cost. For all panels, $\lambda_1=e^{-4.19}$, $\lambda_2=e^{-2.16}$, $\lambda_3=e^{-.042}$ and $\lambda_4=e^{1.31}$.

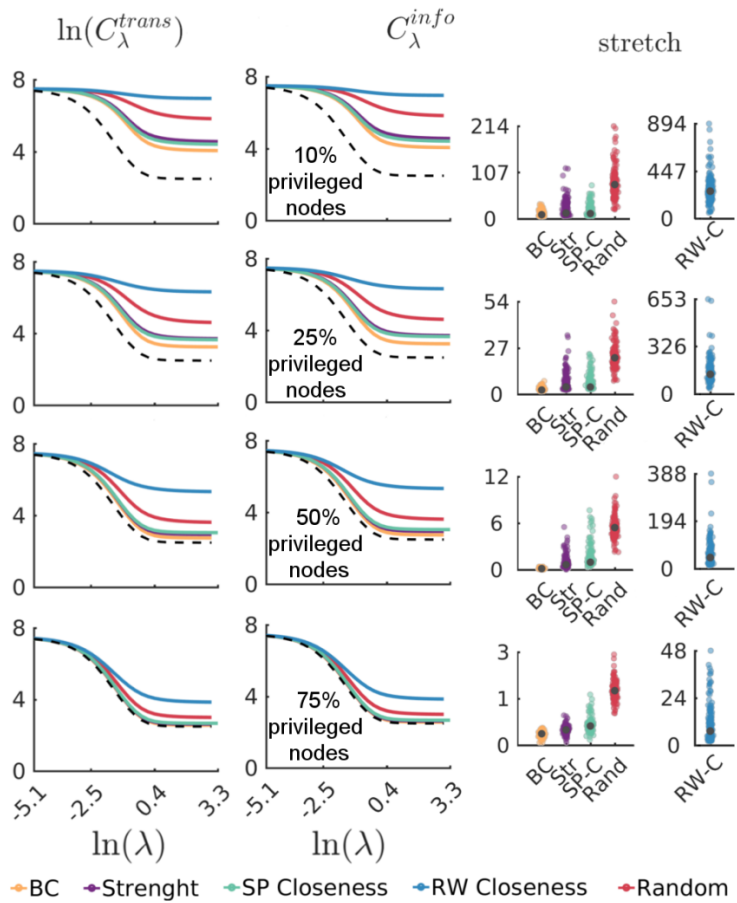

**Fig S10.** Network average values of $C_\lambda^{trans}$ **(left panel)** and $C_\lambda^{info}$ **(middle panel)** as a function of $\lambda$ (node medians across all subjects) for 22, 55, 110, and 165 *privileged nodes* (corresponding to 10%, 25%, 50% and 75% or the network's nodes) that are selected according to betweenness centrality ranking (yellow line), strength

ranking (purple line), shortest-path-based closeness centrality (green line), and random-walk-based closeness centrality (blue line). For comparison purposes, we also show cost measures for randomly sampled nodes (red line). The dotted lines show $C_\lambda^{trans}$ and $C_\lambda^{info}$, respectively, for the case in which all nodes' routing strategies are biased (i.e. 100% privileged nodes). Right panel shows node stretch distributions for the different sets of privileged nodes and centrality rankings. Black markers indicate the median of the distributions.

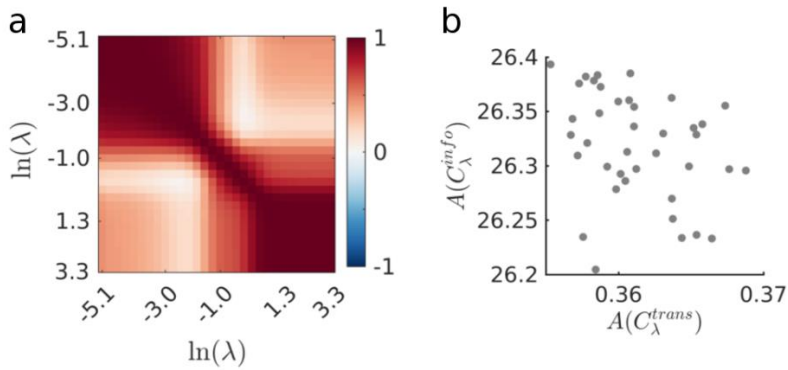

**Fig S11. Communication cost trade-off within subjects.** (a) Correlations between all subject's $C_\lambda^{trans}$ across all values of λ. Positive correlations are colored in red, negative correlations are colored in blue. (b) Scatter plot of the computed areas under the normalized $C_\lambda^{trans}$ and $C_\lambda^{info}$ curves, showing a trade-off between the decay of $C_\lambda^{trans}$ and the growth of $C_\lambda^{info}$ (correlation between A($C_\lambda^{trans}$) and A($C_\lambda^{info}$) is *r = -0.6, p < 0.001)*.